\documentclass[letterpaper, submit]{AAS}			% to submit to JAS

\usepackage{bm}
\usepackage{amsmath, amssymb}
\usepackage[colorlinks=true, pdfstartview=FitV, linkcolor=black, citecolor= black, urlcolor= black]{hyperref}
\usepackage{overcite}
\usepackage{footnpag}			      	% make footnote symbols restart on each page

\usepackage[utf8]{inputenc}
\usepackage{graphicx}
\usepackage[version=4]{mhchem}
\usepackage{siunitx}
\usepackage{subcaption}
\usepackage{adjustbox}
\usepackage{longtable,tabularx}
\usepackage{booktabs}
\usepackage{blkarray}

\PaperNumber{XX-XXX}

\newcommand{\pd}[2]{\frac{\partial #1}{\partial #2}}
\DeclareMathOperator*{\argmin}{arg\,min}

\newcommand{\orcid}[1]{\href{https://orcid.org/#1}{\includegraphics[scale=.012]{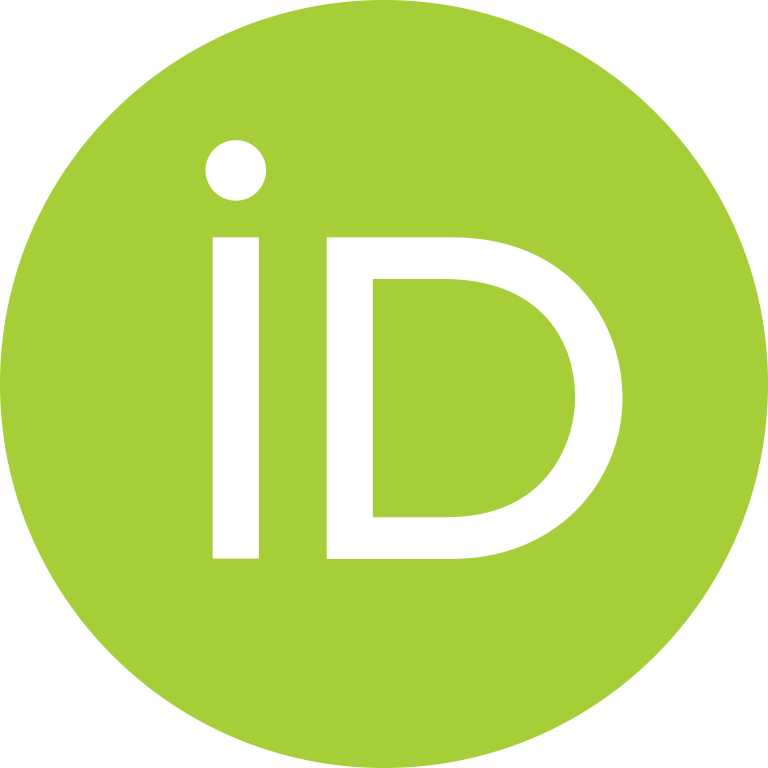}}}

\begin{document}

\title{
%Combining Gradient‑Informed Markov Chain Monte Carlo and Diffusion Models for Indirect Low‑Thrust Trajectory Design
Transfer Learning of Multiobjective Indirect Low-Thrust Trajectories Using Diffusion Models and Markov Chain Monte Carlo
\thanks{
This paper extends material previously presented at the AIAA SciTech 2026 Forum as \emph{Gradient-Informed Monte Carlo Fine-Tuning of Diffusion Models for Low-Thrust Trajectory Design} and at the AAS/AIAA Astrodynamics Specialist Conference 2025 in Boston, MA as \emph{Self-Supervised Diffusion Model Fine-Tuning for Costate
Initialization Using Markov Chain Monte Carlo}.
}
}% with the indirect method?

\author{Jannik Graebner \orcid{https://orcid.org/0009-0003-2497-124X}\thanks{PhD Candidate, Department of Mechanical and Aerospace Engineering, Princeton University, NJ, USA.}
\ and Ryne Beeson \orcid{0000-0003-2176-0976}\thanks{Assistant Professor, Department of Mechanical and Aerospace Engineering, Princeton University, NJ, USA.}
}

\maketitle{} 		

%Maybe change title? could include indirect method

\begin{abstract}
Preliminary low-thrust spacecraft mission design is a global search problem characterized by a complex solution landscape, multiple objectives, and numerous local minima.
During this phase, mission parameters are often not yet fully defined, requiring new solutions to be generated at a high cadence across varying parameter values. 
When combined with the indirect approach to optimal control, diffusion models can accelerate this search by learning distributions that represent high-quality initial costates. 
However, generating training data remains expensive, and opportunities exist to better exploit past data. 
We propose a transfer-learning framework that combines homotopy in a mission parameter with Markov chain Monte Carlo (MCMC) to generate training data more efficiently. 
The approach reformulates a multiobjective optimization problem as sampling from an unnormalized target distribution in costate space.
We compare three MCMC algorithms on a planar multi-revolution transfer in the circular restricted three-body problem, with homotopy in the system mass parameter.
The results show that gradient-based MCMC variants achieve the best trade-off between sample quality and computational cost.
For the test transfer, the proposed framework generates $\SI{40}{\percent}$ more feasible solutions and achieves a higher-quality Pareto front than a state-of-the-art indirect approach based on adjoint control transformations and gradient-based optimization. 
Finally, the MCMC-generated samples are used to fine-tune a diffusion model conditioned on the mass parameter, enabling it to learn a global representation of the underlying solution distribution and efficiently generate new solutions. 
These findings establish the transfer-learning framework as a practical method for efficiently solving indirect trajectory optimization problems with varying parameters.
\end{abstract}


\section{Introduction}
Spacecraft trajectories that minimize both fuel consumption and time of flight are desirable for mission design, but these objectives are typically in conflict with each other. 
It is therefore of interest to mission designers to understand how to efficiently and thoroughly identify Pareto-optimal solutions with respect to these objectives. 
% for long-duration, low-thrust (LT) spacecraft trajectories is challenging and computationally expensive. 
% A high-dimensional solution space, coupled with the nonlinear and non-convex dynamics of the Circular Restricted Three-Body Problem (CR3BP) leads to an optimal control problem with a complex objective landscape and numerous local minima. 
Uncovering the Pareto-optimal solutions can quickly become challenging for low-thrust (LT) transfers in dynamical systems with multiple gravitational bodies, such as the Circular Restricted Three-Body Problem (CR3BP). 
Long-duration LT transfers in such systems are sensitive not only to the nonlinearity of the dynamics, but also to the chaotic behavior of the system. 
During preliminary mission design, in particular, solutions for these transfers must often be generated at a high cadence while the problem parameters are not yet fully defined.
Therefore, the task becomes a global search problem under varying mission parameters.

The first choice that must be made in solving optimal control problems 
% that do not have analytic solutions 
is whether to use a direct or an indirect approach~\cite{Betts.1998}. 
Direct methods transcribe the optimal control problem as stated into a higher-dimensional parameter optimization problem. 
Two benefits of direct methods are that they often have a larger radius of convergence based on initial guesses for the parameter optimization solvers and provide a greater opportunity for endowing physical intuition in the selection of initial guesses. 
The main drawback of direct methods is that they lead to higher-dimensional problems, where the success of the approach is tied to this dimensionality. 
In contrast, the indirect approach maintains the same dimensionality regardless of variations to a base optimal control problem.
It is based on the introduction of the costate variables associated with the first-order necessary conditions for optimality. 
% Applying an indirect optimal control approach reduces the dimension of the solution space through the introduction of costate variables, which become the decision variables of the problem. 
However, the lack of physical intuition for the costates and the high sensitivity of the solution to their values make the practical use of this method challenging. 
Efficient performance therefore depends strongly on generating good initial costate guesses.

\begin{figure}[t!]
\centering
\includegraphics[width=0.8\textwidth]{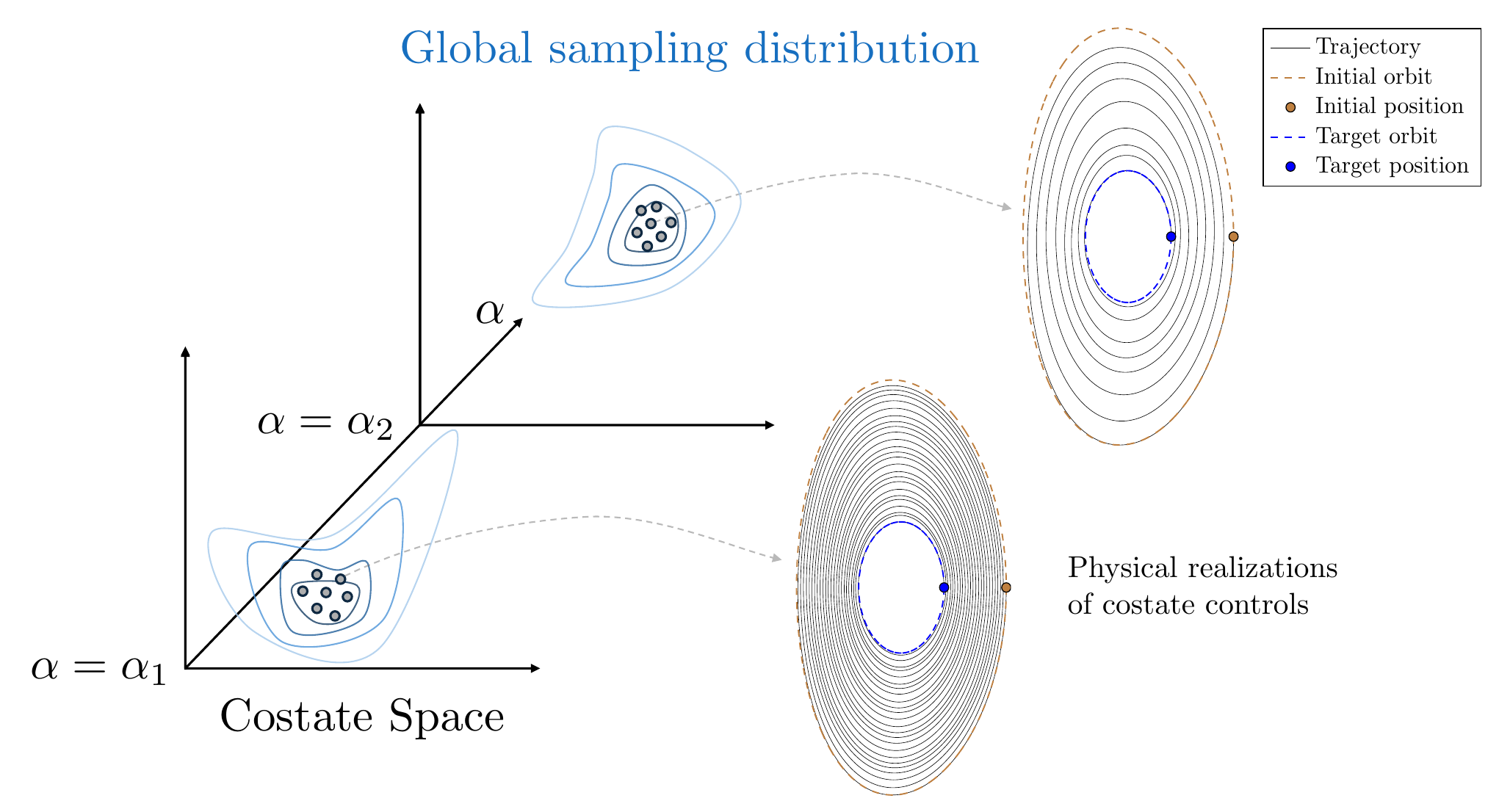}
\caption{Schematic illustration of the global search problem for indirect LT trajectory optimization under a varying mission parameter $\alpha$. 
Clusters in costate space correspond to families of locally optimal trajectories, shown here for an example planar DRO-to-DRO transfer in the Jupiter–Europa system. 
Learning a conditional sampling distribution over these cluster enables efficient generation of new trajectory candidates across parameter values.}
\label{fig: sampling_trajectory_graphic}
\end{figure}

Previous work has shown empirically that locally optimal solutions for trajectory optimization problems form parameter-dependent clusters~\cite{Yam.2011, beeson2024globalsearchoptimalspacecraft, graebner_JAS}.
% However, many existing approaches do not explicitly exploit previously computed solution data when generating new solutions.
It is therefore natural to exploit these structures when searching for new solutions, as was done in our previous work to construct high-quality initial costate guesses~\cite{graebner_JAS}.
% We previously proposed that this structure can be leveraged to construct high-quality initial costate guesses~\cite{graebner_JAS}.
By defining a probability density function supported on solution clusters, the global search problem is recast as sampling from a distribution with unnormalized density. 
This is visualized in Figure~\ref{fig: sampling_trajectory_graphic}, which shows samples in costate space and corresponding trajectory realizations for a distant retrograde orbit (DRO)-to-DRO transfer in the Jupiter-Europa system.
The sample distribution depends on a mission parameter $\alpha$, which could, for example, be the spacecraft's maximum thrust magnitude.
In practice, these solution distributions are defined on a higher-dimensional space and are generally complex, multimodal distributions, making standard parametric approximations inadequate.
This makes diffusion models, which are deep generative machine learning models designed to learn complex data distributions~\cite{SohlDickstein.3122015,Ho.6192020}, a natural choice for accelerating the generation of new solutions.

%Using generative machine learning and MCMC methods, our proposed framework combines two popular families of sampling methods, leveraging the complementary strengths of both approaches.
\begin{figure}[b!]
\centering
\includegraphics[width=1.0\textwidth]{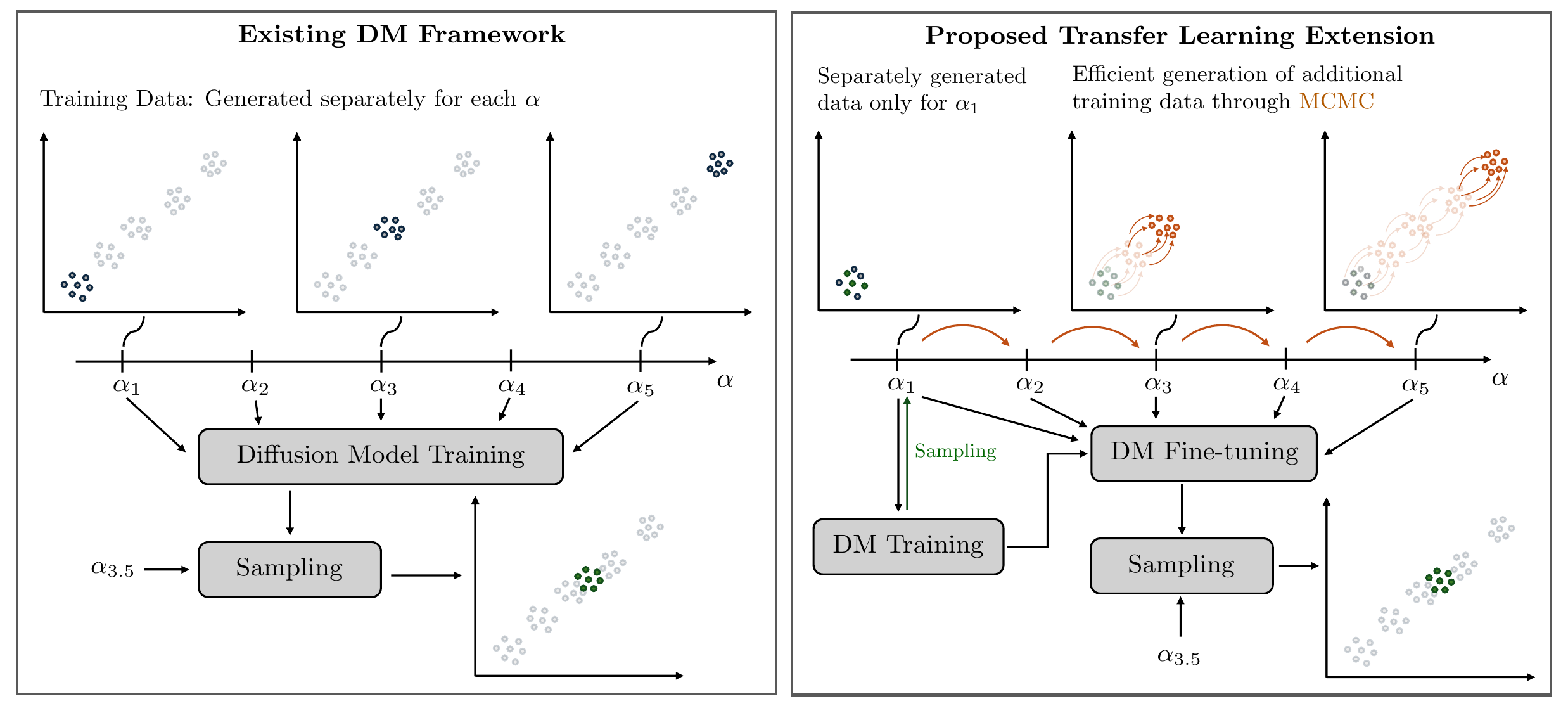}
\caption{Comparison of the previous diffusion model framework~\cite{graebner_JAS} (left) and the proposed transfer-learning extension (right). 
The illustration is schematic; in the actual indirect trajectory-design problem, the samples for each mission parameter $\alpha$ correspond to multimodal, non-Gaussian distributions in costate space.}
\label{fig: transfer_learning_graphic}
\end{figure}

In our previous work, we trained such a model to learn costate distributions for LT transfers and used it to generate high-quality initial guesses for a numerical solver~\cite{graebner_JAS}.
Although this reduced solution-generation time by one to two orders of magnitude compared to classical methods, a key limitation was the high cost of generating the training data. 
In the earlier framework, training data was generated at fixed parameter values within a prescribed range using a gradient-based numerical solver, so that inference effectively relied on interpolation between parameter values already represented in the dataset. 
The left panel of Figure~\ref{fig: transfer_learning_graphic} illustrates this setup schematically: separate datasets are generated for multiple values of $\alpha$, and the diffusion model learns to interpolate between them.

The transfer-learning framework proposed in this paper addresses the limitation of the previous approach by enabling efficient training data generation beyond the parameter values already represented in the original dataset, thereby allowing extrapolation to new mission parameters. 
This is achieved by combining parameter homotopy with MCMC to leverage existing training data when constructing solutions for new parameter values. 

Performing homotopy over full solution distributions, rather than over individual solutions, allows data to be generated across a continuous range of mission-parameter values while keeping successive problems sufficiently similar.
Within this setting, MCMC methods are better suited than gradient-based optimization solvers because they naturally support this type of homotopy over distributions by evolving multiple Markov chains in parallel and avoiding the issue of terminating homotopy paths. 
MCMC is also an appropriate method to deploy due to theoretical results that guarantee asymptotic convergence of the chains to the target distribution~\cite{Douc:2018.springer.mc}. 

At each homotopy step, samples obtained for one parameter value are used to initialize the Markov chains for the next, thereby transferring the learned solution structure across the mission-parameter space.
This idea is illustrated schematically in the right panel of Figure~\ref{fig: transfer_learning_graphic}, where training data is assumed to be available initially only for the parameter value $\alpha_1$, while data for subsequent parameter values is generated through the homotopy-MCMC procedure and then used to fine-tune the diffusion model.

\section{Related Work}
\subsection{Multiobjective Trajectory Optimization}
Various methods have been applied to the multiobjective optimization of LT transfers with the goal of generating Pareto-optimal solutions. 
One straightforward strategy is to use a single-objective optimal control formulation while promoting global exploration in the second objective through randomized initial guesses. 
Russell~\cite{Russell.2007} follows this approach to generate a dense Pareto front for planar transfers in the CR3BP using the indirect method.
Other approaches explicitly address the multiobjective nature of the problem. 
A common choice is evolutionary algorithms such as the non-dominated sorting genetic algorithm (NSGA)~\cite{10.1162/evco.1994.2.3.221} and its extension, NSGA-II~\cite{996017}. 
These algorithms operate on populations of candidate solutions that evolve through selection, crossover, and mutation. 
Similarly to our baseline RWM, they rely only on zeroth-order information about the objective function.
They are frequently combined with gradient-based solvers that refine samples locally, using both direct~\cite{vavrina2009multiobjective} and indirect methods~\cite{COVERSTONECARROLL2000387, Hartmann}. 
Alternative strategies approximate the Pareto set through iterative subdivision and pruning of the design space~\cite{dellnitz, Schuetze01022009}.
More recently, Sullivan et al.~\cite{doi:10.2514/1.A35463} used multiobjective reinforcement learning, specifically Multi-Reward Proximal Policy Optimization, to recover distinct regions of the LT transfer design space in a multi-body setting and employed transfer-learning to initialize training in a higher-fidelity ephemeris model.

All of these approaches typically require substantial computational effort to generate high-quality Pareto fronts. 
Here, high quality refers to fronts that are dense, close to the true Pareto-optimal set, and evenly spaced representation across the objective trade space.
This computational effort is exacerbated for repeated Pareto-front generation under varying mission parameters, since these methods typically do not exploit previously computed solution data to generate new solutions.
This is the key limitation addressed in the present work.

\subsection{Global Search and Data-Driven Methods for Trajectory Optimization}
The idea of exploiting the clustering structure of locally optimal solutions has long been employed in global search algorithms such as Monotonic Basin Hopping (MBH)~\cite{Wales.1997}. 
However, MBH relies on simple sampling distributions and requires extensive manual or automated parameter tuning~\cite{Englander2021}. 
Li et al.~\cite{li2023amortizedglobalsearchtrajectory}
%\textcolor{red}{(fix this sentence as Li et al. is first, also DM not used in these cited refs)}
were the first to leverage generative models, specifically conditional variational autoencoders, to create informed initial guesses for trajectory optimization. 
In subsequent work, they formalized and extended this approach and showed that diffusion models further improve initial guess generation~\cite{beeson2024globalsearchoptimalspacecraft,Li:2024.iclr.05571v4}.
Trained on databases of pre-computed solutions across multiple thrust levels, their models warm-started a direct optimization solver for unseen thrust values, substantially improving convergence rates.
The same framework was later used to condition on other mission parameters, such as boundary conditions and objective function parameters~\cite{graebner2024learningoptimalcontrol}.
Extending these ideas, we proposed a global-search framework that pairs diffusion models with the indirect optimal-control method for LT trajectory design~\cite{graebner_JAS}. 
The present work extends this line of research with a focus on generating new training data more efficiently.

In guidance problems about a nominal solution, there is a natural approach based on backpropagating costates that satisfy the transversality conditions to generate large training datasets for deep neural networks in an efficient manner~\cite{doi:10.2514/1.G005254}. 
The work in this manuscript goes beyond the aforementioned problem setting, as we are interested in uncovering many qualitatively different nominal solutions, and the parameter variations of our problem need not be simple boundary condition variations. 

\subsection{Homotopy Approaches for Trajectory Optimization}
Homotopy and continuation methods are well-established tools in trajectory optimization and optimal control. 
In LT trajectory design, they are commonly used to improve convergence by deforming an easier auxiliary problem into the target problem.
Examples include continuation in the objective, thrust magnitude, or other problem parameters \cite{doi:10.2514/1.4022,ZHANG2023110798}. 
Related smoothing and continuation ideas have also been used to regularize bang-bang problems in optimal control and improve the robustness of indirect shooting methods~\cite{Bertrand.2002,Taheri.2016}. 
More elaborate homotopy schemes, including double homotopy, two-phase homotopy, and smoothing-kernel approaches, have been shown to further improve the convergence of indirect optimization solvers~\cite{doi:10.2514/1.G001553,doi:10.2514/1.A34144,doi:10.2514/1.G006977}. 
In these settings, homotopy is typically used to continue individual solutions or families of extremals as the problem is gradually transformed.

Our work differs from the approaches above in two distinct ways.
First, rather than performing homotopy on individual solutions, we apply it to full solution distributions by continuing samples that are representative of those distributions. 
Second, our primary motivation for introducing homotopy is not restricted to transitioning from an easier problem to a more difficult one, but to the generation of training data across continuously varying values of a mission parameter. 
Although the test case considered in this work follows such a progression, the proposed framework is not restricted to that setting.
    
\section{Contributions}

Three main contributions are made in this manuscript. 
We briefly state them now and provide further details in the paragraphs that follow:
\begin{itemize}
    \item Introduction of a novel transfer-learning approach for multiobjective indirect trajectory optimization.
    % \item Development of an efficient MCMC formulation to sample high-quality costates.
    \item Development of efficient MCMC formulations that use an advanced preliminary screening algorithm and faithful surrogate gradients to sample high-quality costates. 
    % \item Development of efficient problem evaluation using an advanced preliminary screening algorithm and MCMC formulations that use a faithful surrogate gradient function to sample high-quality costates.
    \item Integration of the transfer-learning approach into a self-supervised fine-tuning framework for diffusion models.
\end{itemize}
% These contributions are described in more detail below.

The primary contribution of this work is the introduction of a novel transfer-learning approach that combines MCMC and parameter homotopy for multiobjective indirect trajectory optimization. 
Starting from existing solutions for one parameter value, this new framework enables the efficient and automated generation of new training data for diffusion models.
Equivalently, the proposed framework can be viewed as a method for extrapolating across a family of conditional distributions associated with solutions for different mission-parameter values.
%The framework relies on a novel objective-function formulation that combines multiple objectives, which can be weighted individually to explicitly target specific regions of the Pareto front. 
%Based on this objective formulation, we define a parameter-dependent target distribution with unnormalized density, which provides the basis for the MCMC sampling procedure.

A second contribution is the development of an efficient MCMC formulation to sample high-quality costates within this framework.
At its core, the formulation defines a parameter-dependent target distribution whose unnormalized density depends on an objective function that combines constraint violation, fuel consumption, and time of flight.
The relative weights of these terms can be adjusted to explicitly target specific regions of the Pareto front.
To sample from this distribution, we implement and compare three MCMC variants: random-walk Metropolis (RWM)~\cite{Metropolis.1953}, which serves as a baseline, the Metropolis-adjusted Langevin algorithm (MALA)~\cite{Roberts.1998}, and Hamiltonian Monte Carlo (HMC)~\cite{Neal.2012}.
The objective function is evaluated efficiently through a preliminary screening algorithm at each MCMC iteration.
For the gradient-based variants, MALA and HMC, we show that the gradient can be approximated effectively through a fixed-time objective.
This surrogate gradient is computed using analytic derivatives obtained from the numerical propagation of state transition matrices (STMs).

A third contribution is the integration of the transfer-learning approach into a self-supervised diffusion-model fine-tuning framework for learning a global solution distribution across parameter values. 
MCMC-generated samples for different values of a mission parameter are used to fine-tune a diffusion model so that it can generate new solutions for previously unseen or sparsely sampled parameter values. 
For fine-tuning, we employ a reward-weighted loss function based on reward values assigned to samples from the final MCMC iterations. 
These rewards place greater emphasis on high-quality samples, that is, samples with lower objective values. 
By using generated samples to warm-start a numerical solver, the fine-tuned diffusion model can then rapidly generate new solutions, effectively producing a dense Pareto front for the problem with new parameters.

\section{Method Validation}
We validate the proposed framework on a planar multi-revolution DRO transfer with homotopy in the CR3BP mass parameter. 
Starting from solutions in the Jupiter-Europa system, the homotopy-MCMC procedure generates solutions for analogous transfers in the Saturn-Titan system. 
A homotopy in the mass parameter provides a challenging test case that illustrates the broader applicability of the framework to other mission parameters, such as thrust magnitude or boundary conditions. 

One might also understand the parameter variation problem considered in our validation work as motivated by a situation where an extensive prior design in the Jupiter-Europa system had been completed and at some future time a new design in the Saturn-Titan system is needed. 
The benefit of an efficient transfer-learning approach, as proposed in this work, is that intermediate solutions will also be generated and that it provides the ability to accelerate yet undefined design needs. 

A comparison of the three MCMC algorithms shows that incorporating gradient information from the target distribution improves both sampling efficiency and sample quality. 
Among the methods considered, MALA achieves the highest sample quality for comparable runtime. 
It also outperforms a state-of-the-art approach that initializes costates in a gradient-based solver using adjoint control transformations (ACT) in both solution quality and the number of solutions found for a fixed runtime.

\section{Organization of this Work}
We begin with the \emph{\nameref{sec: problem formulation}}, which presents the indirect approach to spacecraft trajectory optimization. 
This includes an introduction to the general \emph{\nameref{sec: optimal control problem}}, the specific formulation of \emph{\nameref{sec: low-thrust trajectory optimization}}, and the application of \emph{\nameref{sec: primer vector}}. 
This section also presents \emph{\nameref{sec: AD}} for the indirect method and provides a brief description of the \hyperref[sec: CR3BP]{\emph{CR3BP}}. 
We then introduce our novel contributions in the \emph{\nameref{sec: methodology}} section, starting with an \emph{\nameref{sec: proposed framework}}. 
This is followed by a detailed description of the individual components of the framework: \emph{\nameref{sec: diffusion models}}, \emph{\nameref{sec: fine-tuning}}, \emph{\nameref{sec: gradient-based MCMC}}, \emph{\nameref{sec: objective function}}, and \emph{\nameref{sec: gradient approximation}}. 
Finally, we present the \emph{\nameref{sec: results}}, discuss the \emph{\nameref{sec: limitations}}, and give a \emph{\nameref{sec: conclusion}}.

\section{Problem Formulation}
\label{sec: problem formulation}

The next three subsections on the \emph{\nameref{sec: optimal control problem}}, \emph{\nameref{sec: low-thrust trajectory optimization}}, and \emph{\nameref{sec: primer vector}} follow the definitions and setup of our previous work~\cite{graebner_JAS}, but we include these sections here for the convenience of the reader and to establish a uniform notation for the remainder of the manuscript. 
% Parts of the following setup are similar to our previous work , and we restate the problem setup for the convenience of the reader.

\subsection{Optimal Control Problem}
\label{sec: optimal control problem}
Optimal control aims to determine a feasible control history $\boldsymbol{u}(t)$ for a dynamical system that minimizes a specified performance measure while satisfying state constraints~\cite{Kirk.2004}.
A general version of the problem is given by
\begin{align}
    \label{equation: cost functional}
    \min_{\boldsymbol{u},\,t_f} \left\{ J(\boldsymbol{u}) \equiv \phi(\boldsymbol{x}(t_f), t_f) + \int_{t_0}^{t_f} \mathcal{L}(\boldsymbol{x}(t), \boldsymbol{u}(t), t) \, dt \right\} 
    \quad 
    \textrm{subject to Eqs.} \ 
    \eqref{equation: evolution differential equation}, \eqref{equation: initial and final boundary conditions}, \eqref{equation: path constraints}, 
\end{align}
where $J$ is the objective function that includes a running cost $\mathcal{L}$ and a terminal cost $\phi$.
We consider a problem with free final time over the interval $t\in[t_0,t_f]$.
The state $\boldsymbol{x}(t)$ evolves according to the dynamics of the system
\begin{align}
    \label{equation: evolution differential equation}
    \dot{\boldsymbol{x}}(t) = \boldsymbol{f}(\boldsymbol{x}(t), \boldsymbol{u}(t), t), \quad \forall t \in [t_0, t_f],
\end{align}
and satisfies the initial and terminal boundary conditions
\begin{align}
    \label{equation: initial and final boundary conditions}
    \boldsymbol{x}(t_0) = \boldsymbol{x}_0, \quad  \boldsymbol{\psi}\left[ \boldsymbol{x}(t_f), t_f \right] = \boldsymbol{0}.
\end{align}
Both the natural dynamics and control-induced accelerations are included in the vector field $\boldsymbol{f}$.
Additional constraints on the state and control are prescribed as equality path constraints
\begin{align}
    \label{equation: path constraints}
    \boldsymbol{\xi}(\boldsymbol{x}(t), \boldsymbol{u}(t), t) = \boldsymbol{0}, \quad & \forall t \in [t_0, t_f].
\end{align}
Inequality path constraints will not be considered in this work, but could be included.

\subsection{Low-Thrust Spacecraft Trajectory Optimization}
\label{sec: low-thrust trajectory optimization}
%When optimizing the trajectory of a spacecraft with low thrust propulsion, a primary goal is generally to minimize fuel consumption.
%We therefore choose the objective function in Mayer form as
%\begin{equation}
%	\label{eq:cost function}
%	J = -km_f,
%\end{equation}
%where $m_f$ denotes the final spacecraft mass and $k>0$ provides an additional degree of freedom.
%The secondary objective of minimizing transfer time is indirectly imposed through a maximum time threshold and a thorough exploration of the solution space.
Our goal is to find Pareto-optimal trajectories for an LT transfer, minimizing fuel consumption and time of flight simultaneously. 
A trajectory is labeled Pareto-optimal if no other feasible solution attains strictly lower values for both objectives.
Only the fuel consumption appears explicitly in the cost function as $J=-km_f$, where $k>0$ is a fixed scaling parameter and $m_f$ is the final mass. 
The secondary objective of minimizing time of flight is not directly included in the optimal control problem, but later targeted through the MCMC algorithm.
The propulsion system has maximum thrust $T_\mathrm{max}$ and constant specific impulse $I_\mathrm{sp}$ with exhaust velocity $c = I_\mathrm{sp} g_0$, where $g_0$ is the standard gravitational acceleration.
The state vector $\boldsymbol{x}=(\boldsymbol{r}^\top,\boldsymbol{v}^\top, m) \in \mathbb{R}^{7}$ includes the spacecraft's position $\boldsymbol{r}\in\mathbb{R}^3$, velocity $\boldsymbol{v}\in\mathbb{R}^3$, and mass $m \in \mathbb{R}_{\ge 0}$.
It evolves according to the autonomous dynamical system
\begin{equation}
\label{Equation: dynamics}
    \dot{\boldsymbol{x}} = \boldsymbol{f}(\boldsymbol{x},\boldsymbol{u}) =
	\begin{pmatrix}
	\dot{\boldsymbol{r}} \\
	\dot{\boldsymbol{v}} \\
	\dot{m}
	\end{pmatrix}
	=
	\begin{pmatrix}
	\boldsymbol{v} \\
	\boldsymbol{g}(\boldsymbol{r},\boldsymbol{v}) + \frac{T}{m}\hat{\boldsymbol{u}} \\
	-\frac{T}{c}
	\end{pmatrix},
\end{equation}
where the natural acceleration of the system is described by the vector field $\boldsymbol{g}$.
The control vector of the system consists of three components: $\boldsymbol{u}=(\hat{\boldsymbol{u}}^\top,T,\zeta)^\top$.
The thrust direction is described by the unit vector $\hat{\boldsymbol{u}}\in\mathbb{R}^3$ and its magnitude is controlled through the throttle $\sigma$, where $T=\sigma T_\mathrm{max}$.
To avoid writing the bounds on $\sigma \in [0,1]$ as inequality constraints, a slack variable $\zeta$ with $\sigma=\sin^2 \zeta$ is introduced.
This allows us to express the set of admissible controls as two equality path constraints:
\begin{equation}
    \boldsymbol{\xi}(\boldsymbol{x}(t), \boldsymbol{u}(t), t) = 
    \begin{pmatrix}
	\hat{\boldsymbol{u}}^\top\hat{\boldsymbol{u}} - 1 \\
    T-T_\mathrm{max}\sin^2\zeta
    \end{pmatrix} = \boldsymbol{0}.
\end{equation}
The trajectory starts from a fixed initial state $\boldsymbol{x}_0$ and targets a fixed terminal position $\boldsymbol{r}_f$ and velocity $\boldsymbol{v}_f$:
\begin{equation}
\label{Equation: Boundary conditions trajectory}
	\boldsymbol{x}(t_0) = \boldsymbol{x}_0, \quad\boldsymbol{e}(\boldsymbol{x}(t_f)) =
    \begin{pmatrix}
	\boldsymbol{r}_f - \boldsymbol{r}(t_f) \\
    \boldsymbol{v}_f - \boldsymbol{v}(t_f) 
    \end{pmatrix} = \boldsymbol{0},
\end{equation}
where $\boldsymbol{e}$ denotes the constraint violation vector. 
The final time $t_f$ is a free variable.

\subsection{Primer Vector Theory}
\label{sec: primer vector}
Primer Vector Theory applies an indirect approach to the optimal control problem defined in Eq.~\eqref{equation: cost functional}, in the context of spacecraft trajectory optimization~\cite{Lawden1964OptimalTF}.
Using techniques from calculus of variations, the first‑order necessary conditions for optimality are applied to reformulate Eq.~\eqref{equation: cost functional} as a two-point boundary value problem~\cite{Liberzon.2012}.  
The dimension of the state vector is doubled through the introduction of a costate vector $\boldsymbol{\lambda} =(\boldsymbol{\lambda}_r^\top,\boldsymbol{\lambda}_v^\top, \lambda_m)^\top \in \mathbb{R}^{7}$, consisting of position, velocity, and mass costates.
The Hamiltonian of the system is given by 
\begin{equation}
	\label{Equation: Problem Hamiltonian}
	H = \mathcal{L} + \boldsymbol{\lambda}^\top \boldsymbol{f}(\boldsymbol{x},\boldsymbol{u}) = \boldsymbol{\lambda}_r^\top \boldsymbol{v} + \boldsymbol{\lambda}_v^\top \left(\boldsymbol{g}(\boldsymbol{r},\boldsymbol{v}) + \frac{T}{m}\hat{\boldsymbol{u}}\right) - \lambda_m\frac{T}{c}.
\end{equation}
According to Pontryagin's Minimum Principle~\cite{Pontryagin.2018}, we set $\hat{\boldsymbol{u}} = -\boldsymbol{\lambda}_v/\lambda_v$, to minimize $H$.
Throughout this paper, writing a vector without boldface (e.g., $\lambda_v$) denotes its Euclidean norm.
By introducing a switching function $S = \lambda_v + \lambda_m m / c$ we reformulate the Hamiltonian as
\begin{equation}
	\label{Equation: Hamiltonian without control}
	H = \boldsymbol{\lambda}_r^\top \boldsymbol{v} + \boldsymbol{\lambda}_v^\top \boldsymbol{g}(\boldsymbol{r},\boldsymbol{v}) - S\frac{T}{m}.
\end{equation}
To minimize $H$, the throttle is chosen based on the sign of $S$, resulting in the discontinuous bang-bang control law~\cite{Conway.2010}:
\begingroup
\small
\renewcommand{\arraystretch}{0.9}
\begin{equation}
\label{Equation: Control Law}
\hat{\boldsymbol{u}}
=
-\frac{\boldsymbol{\lambda}_v}{\lambda_v},
\qquad
\sigma =
\begin{cases}
0 & \text{if } S < 0, \\
1 & \text{if } S > 0, \\
0 \leq \sigma \leq 1 & \text{if } S = 0.
\end{cases}
\end{equation}
\endgroup
Applying the first-order necessary conditions for optimality yields the costate dynamics~\cite{Liberzon.2012}
\begingroup
\small
\renewcommand{\arraystretch}{0.9}
\begin{equation}
\label{Equation: Costate equation of motion}
\dot{\boldsymbol{\lambda}}
=
\begin{pmatrix}
\dot{\boldsymbol{\lambda}}_r \\
\dot{\boldsymbol{\lambda}}_v \\
\dot{\lambda}_m
\end{pmatrix}
=
\begin{pmatrix}
-\boldsymbol{G}^\top \boldsymbol{\lambda}_v \\
-\boldsymbol{\lambda}_r - \boldsymbol{H}^\top \boldsymbol{\lambda}_v \\
-\lambda_v T/m^2
\end{pmatrix},
\quad
\text{where}
\quad
\boldsymbol{G}=\frac{\partial \boldsymbol{g}}{\partial \boldsymbol{r}},
\qquad
\boldsymbol{H}=\frac{\partial \boldsymbol{g}}{\partial \boldsymbol{v}}.
\end{equation}
\endgroup
The equations of motion for the combined state and costate vector $\boldsymbol{y}\in \mathbb{R}^{14}$ are therefore given by
\begingroup
\small
\renewcommand{\arraystretch}{0.9}
\begin{equation}
\label{Equation: Combined state equations}
\dot{\boldsymbol{y}}=\boldsymbol{f}(\boldsymbol{y}) = 
\begin{pmatrix}
\dot{\boldsymbol{r}} \\
\dot{\boldsymbol{v}} \\
\dot{m} \\
\dot{\boldsymbol{\lambda}}_r \\
\dot{\boldsymbol{\lambda}}_v \\
\dot{\lambda}_m
\end{pmatrix}
=
\begin{pmatrix}
\boldsymbol{v} \\
\boldsymbol{g}(\boldsymbol{r},\boldsymbol{v}) - (\boldsymbol{\lambda}_v/\lambda_v)T/m \\
-T/c \\
-\boldsymbol{G}^\top \boldsymbol{\lambda}_v \\
-\boldsymbol{\lambda}_r - \boldsymbol{H}^\top \boldsymbol{\lambda}_v \\
-\lambda_v T/m^2
\end{pmatrix}.
\end{equation}
\endgroup
Combined with~\eqref{Equation: Boundary conditions trajectory} and Eq.~\eqref{Equation: Control Law} this yields a two-point boundary value problem.
Since the final time and final mass are free, the transversality conditions are
\begin{equation}
\label{Equation: lam_m=-k}
\lambda_m(t_f)
= \left.\frac{\partial \phi}{\partial m}\right|_{m=m_f}=-k
\quad \text{and} \quad
H(\boldsymbol{x}(t_f), \boldsymbol{u}(t_f), \boldsymbol{\lambda}(t_f)) = 0.
\end{equation}
We choose $k$ such that $\lambda_m(t_0)=-1$, removing the initial mass costate from the search space.
The problem is relaxed by not enforcing the condition on the Hamiltonian explicitly and instead implicitly targeting it through the global search.

\subsection{Analytic Derivatives}
\label{sec: AD}
Employing a gradient-based method to find feasible solutions of the two-point boundary-value problem requires the partial derivatives of the constraint violations and fuel consumption with respect to the costates at the initial time.
These derivatives are obtained by numerically propagating the state transition matrix (STM), defined as $\boldsymbol{\Phi}(t,t_0) = \pd{\boldsymbol{y}(t)}{\boldsymbol{y}(t_0)}$.
The STM is governed by the matrix differential equation 
\begin{equation}
\label{eq: MDE}
    \dot{\boldsymbol{\Phi}}(t,t_0) = \boldsymbol{A}(t)\boldsymbol{\Phi}(t,t_0), 
\end{equation}
with initial condition $\boldsymbol{\Phi}(t_0,t_0)=\boldsymbol{I}$.
Here, the Jacobian $\boldsymbol{A}(t)\in\mathbb{R}^{14\times 14}$ of the system in Eq.~\eqref{Equation: Combined state equations} is given by
\begingroup
\small
\setlength{\arraycolsep}{3pt}
\renewcommand{\arraystretch}{0.9}
\begin{equation}
\label{eq: State Jacobian}
\boldsymbol A(t)=\frac{\partial \boldsymbol{f}}{\partial \boldsymbol{y}} =
\left(
\begin{array}{@{}cccccc@{}}
\boldsymbol{0} & \boldsymbol{I} & \boldsymbol{0} & \boldsymbol{0} & \boldsymbol{0} & \boldsymbol{0} \\
\boldsymbol{G} & \boldsymbol{H} &
\dfrac{\boldsymbol{\lambda}_v}{\lambda_v}\dfrac{T}{m^2} &
\boldsymbol{0} &
-\dfrac{T}{m}\!\left(\dfrac{\boldsymbol{I}}{\lambda_v} -
\dfrac{\boldsymbol{\lambda}_v \boldsymbol{\lambda}_v^\top}{\lambda_v^3}\right) &
\boldsymbol{0} \\
\boldsymbol{0} & \boldsymbol{0} & 0 & \boldsymbol{0} & \boldsymbol{0} & 0 \\
-\dfrac{\partial (\boldsymbol{G}^\top \boldsymbol{\lambda}_v)}{\partial \boldsymbol{r}} &
\boldsymbol{0} & \boldsymbol{0} & \boldsymbol{0} & -\boldsymbol{G}^\top & \boldsymbol{0} \\
\boldsymbol{0} &
-\dfrac{\partial (\boldsymbol{H}^\top \boldsymbol{\lambda}_v)}{\partial \boldsymbol{v}} &
\boldsymbol{0} & -\boldsymbol{I} & -\boldsymbol{H}^\top & \boldsymbol{0} \\
\boldsymbol{0} & \boldsymbol{0} &
\dfrac{2 \lambda_v T}{m^3} &
\boldsymbol{0} &
-\dfrac{\boldsymbol{\lambda}_v^\top T}{\lambda_v m^2} &
0
\end{array}
\right).
\end{equation}
\endgroup
This Jacobian is valid for both thrust and coast arcs by setting $T=T_\mathrm{max}$ and $T=0$, respectively. 
Note, however, that the STM mapping of small perturbations from the initial to final state is only valid on continuous trajectories. 
Switching from a thrust to a coast arc and vice versa introduces a total of $N$ discontinuities between the $N+1$ arcs of the trajectory. 
To determine the sensitivity of the final state with respect to the initial state, we apply the chain rule
\begin{equation}
\frac{\partial \boldsymbol{y}(t_f)}{\partial \boldsymbol{y}(t_0)}
= \boldsymbol{\Phi}(t_f, t_{N+}) \boldsymbol{\Psi}_N \boldsymbol{\Phi}(t_{N-}, t_{(N-1)+}) 
\boldsymbol{\Psi}_{N-1} \cdots 
\boldsymbol{\Phi}(t_{2-}, t_{1+}) \boldsymbol{\Psi}_1 \boldsymbol{\Phi}(t_{1-}, t_0),
\end{equation}
where the mapping across a discontinuity at time $t_n$ is described by the matrix
\begin{equation}
\label{eq: discontinuity}
\boldsymbol{\Psi}_n \equiv 
\frac{\partial \boldsymbol{y}(t_{n+})}{\partial \boldsymbol{y}(t_{n-})}
= \boldsymbol{I}_{14\times 14} 
+ (\dot{\boldsymbol{y}}|_{t_{n+}} - \dot{\boldsymbol{y}}|_{t_{n-}})
\bigl(\tfrac{\partial S}{\partial \boldsymbol{y}} / \dot{S}|_{t_{n-}}\bigr).
\end{equation}
Here we use $t_{n-}$ and $t_{n+}$ to describe the state immediately before and after the discontinuity. 
The additional non-identity term in Eq.~\eqref{eq: discontinuity} originates from a change in the switching time due to a perturbation of the state at $t_{n-}$, which then results in a perturbation of the state at $t_{n+}$. 
A more detailed explanation is provided by Russell~\cite{Russell.2007}.
Evaluating the derivatives yields
\begingroup
\small
\setlength{\arraycolsep}{4pt}
\renewcommand{\arraystretch}{0.95}
\begin{equation}
\boldsymbol{\Psi}_n=\boldsymbol{I}_{14\times 14}
+
\frac{1}{
\hat{\boldsymbol{\lambda}}_v^\top\dot{\boldsymbol{\lambda}}_v|_{t_n}
+
\frac{1}{c}\left[
\dot{m}|_{t_{n-}}\lambda_m
+
\dot{\lambda}_m|_{t_{n-}}m
\right]
}
\left[
\begin{array}{@{}cccccc@{}}
\boldsymbol{0} & \boldsymbol{0} & \boldsymbol{0} & \boldsymbol{0} & \boldsymbol{0} & \boldsymbol{0}\\
\boldsymbol{0} &
\boldsymbol{0} &
\displaystyle \frac{\Delta T\,\lambda_m}{mc}\,\hat{\boldsymbol{\lambda}}_v &
\boldsymbol{0} &
\displaystyle \frac{\Delta T}{m}\,\hat{\boldsymbol{\lambda}}_v\,\hat{\boldsymbol{\lambda}}_v^{\!\top} &
\displaystyle \frac{\Delta T}{c}\,\hat{\boldsymbol{\lambda}}_v
\\[3pt]
\boldsymbol{0} &
\boldsymbol{0} &
\displaystyle \frac{\Delta T\,\lambda_m}{c^2} &
\boldsymbol{0} &
\displaystyle \frac{\Delta T}{c}\,\hat{\boldsymbol{\lambda}}_v^{\!\top} &
\displaystyle \frac{\Delta T\,m}{c^2}
\\[3pt]
\boldsymbol{0} & \boldsymbol{0} & \boldsymbol{0} & \boldsymbol{0} & \boldsymbol{0} & \boldsymbol{0}\\
\boldsymbol{0} & \boldsymbol{0} & \boldsymbol{0} & \boldsymbol{0} & \boldsymbol{0} & \boldsymbol{0}\\
\boldsymbol{0}&
\boldsymbol{0} &
\displaystyle \frac{\Delta T\,\lambda_m\,\lambda_v}{m^{2}c} &
\boldsymbol{0} &
\displaystyle \frac{\Delta T}{m^{2}}\,{\boldsymbol{\lambda}}_v^{\!\top} &
\displaystyle \frac{\Delta T\,\lambda_v}{mc}
\end{array}
\right],
\end{equation}
\endgroup
where $\hat{\boldsymbol{\lambda}}_v=\boldsymbol{\lambda}_v/\lambda_v$ and $m$, $\boldsymbol{\lambda}_v$ and $\lambda_m$ are all evaluated at time $t_n$ and
\begin{equation}
    \Delta T = T(t_{n-})-T(t_{n+})= \begin{cases}
        T_\mathrm{max}, & \text{if switching from thrust arc to coast arc} \\
        -T_\mathrm{max}, & \text{if switching from coast arc to thrust arc.}
    \end{cases}
\end{equation}
The required partial derivatives can then be extracted as the corresponding submatrices from the STM chain
\begin{equation}
\label{eq: STM G1 and G2}
\begin{aligned}
\boldsymbol{G}_1
&= \begin{bmatrix} 
\displaystyle \pd{\boldsymbol{e}}{\boldsymbol{\lambda}_r(t_0)} & \displaystyle \pd{\boldsymbol{e}}{\boldsymbol{\lambda}_v(t_0)} 
\end{bmatrix}=
- \begin{bmatrix}
\displaystyle \frac{\partial \boldsymbol{r}(t_f)}{\partial \boldsymbol{\lambda}_r(t_0)}
&
\displaystyle \frac{\partial \boldsymbol{r}(t_f)}{\partial \boldsymbol{\lambda}_v(t_0)}
\\
\displaystyle \frac{\partial \boldsymbol{v}(t_f)}{\partial \boldsymbol{\lambda}_r(t_0)}
&
\displaystyle \frac{\partial \boldsymbol{v}(t_f)}{\partial \boldsymbol{\lambda}_v(t_0)}
\end{bmatrix}
= 
-\left[
\frac{\partial \boldsymbol{y}(t_f)}{\partial \boldsymbol{y}(t_0)}
\right]_{1:6,\;8:13},
\\
\boldsymbol{G}_2
&=
- \begin{bmatrix}
\displaystyle \frac{\partial m(t_f)}{\partial \boldsymbol{\lambda}_r(t_0)}
&
\displaystyle \frac{\partial m(t_f)}{\partial \boldsymbol{\lambda}_v(t_0)}
\end{bmatrix}
=
-\left[
\frac{\partial \boldsymbol{y}(t_f)}{\partial \boldsymbol{y}(t_0)}
\right]_{7,\;8:13}.
\end{aligned}
\end{equation}
We extract only the derivatives with respect to the position and velocity costates, since the initial mass costate is fixed.

The discontinuity mapping is often neglected, as it only contributes a small change in the STM if the trajectory includes only a small number of switches. 
However, for the multi-revolution transfers in this work, which often include more than $20$ switches, our analysis showed that including this additional term is crucial for accurate derivatives.

\subsection{Circular Restricted Three-Body Problem}
\label{sec: CR3BP}
The CR3BP is a widely used idealization for preliminary trajectory design in astrodynamics. 
In this model, two massive bodies govern the motion of a third body with negligible mass, so that only the gravitational influence of the two massive bodies is considered. 
Let the two massive bodies have masses $m_1$ and $m_2$ with $m_1 > m_2$, referred to as the primary and secondary, respectively.
We adopt the standard system of non-dimensional natural units (NU) in which distances, times, and masses are normalized: the distance unit (DU) equals the separation between the two primaries; the time unit (TU) is their orbital period divided by $2\pi$; and the mass unit (MU) is $m_1+m_2$. 
Under this normalization, the single dimensionless parameter of the problem is the mass ratio $\mu = m_2/(m_1+m_2)$.
The equations of motion are expressed in a uniformly rotating reference frame in which the primaries remain fixed on the $r_1$-axis at locations $r_1=-\mu$ and $r_1 = 1-\mu$, respectively. 
The resulting gravitational field is described by
\begin{equation}
    \label{Equation: CR3BP dynamical system}
    \ddot{\boldsymbol{r}} = 
    \begin{pmatrix}
        \ddot{r}_1 \\
        \ddot{r}_2 \\
        \ddot{r}_3 
    \end{pmatrix}
    = \boldsymbol{g}(\boldsymbol{r},\boldsymbol{v}) =
    \begin{pmatrix}
        2v_2 + r_1 - (1-\mu)\frac{r_1+\mu}{\rho_1^3} - \mu\frac{r_1-1+\mu}{\rho_2^3} \\
        -2v_1 + r_2 - (1-\mu)\frac{r_2}{\rho_1^3} - \mu\frac{r_2}{\rho_2^3} \\
        -(1-\mu)\frac{r_3}{\rho_1^3} - \mu\frac{r_3}{\rho_2^3}
    \end{pmatrix},
\end{equation}
where
$\rho_1 = \sqrt{(r_1+\mu)^2+r_2^2+r_3^2}$ and $\rho_2 = \sqrt{(r_1-1+\mu)^2+r_2^2+r_3^2}$ denote the distances to the primary and secondary, respectively.

\section{Methodology}
\label{sec: methodology}
\subsection{Overview of the Proposed Framework}
\label{sec: proposed framework}
Our goal is to generate high-quality initial costates \(\boldsymbol{\lambda}(t_0)\) for an LT spacecraft transfer.
Each initial costate defines a trajectory realization of the combined state \(\boldsymbol{y}(t)\) when propagated under the dynamics of Eq.~\eqref{Equation: Combined state equations}, with the continuous control \(\boldsymbol{u}(t)\) determined by the control law in Eq.~\eqref{Equation: Control Law}.
To simplify notation in the remainder of this work, \(\boldsymbol{\lambda}\) denotes the costate vector at the initial time, excluding the fixed mass costate.
We define a high-quality costate as one that yields a trajectory \(\boldsymbol{y}(t)\) and control history \(\boldsymbol{u}(t)\) with low constraint violation \(e\), low fuel consumption \(\Delta m\), and short time of flight \(\tau_s\).
These three performance measures are combined into the objective function
\begin{equation}
    \label{eq: objective}
    J^*(\boldsymbol{\lambda};\alpha)
    \equiv
    e^*(\boldsymbol{\lambda};\alpha)
    +
    \kappa_1 \left(
    \frac{\Delta m^*(\boldsymbol{\lambda};\alpha)}{m_0}
    +
    \kappa_2 \tau_s^*(\boldsymbol{\lambda};\alpha)
    \right),
\end{equation}
where \(\alpha\) denotes a mission parameter and the parameters \(\kappa_1,\kappa_2\in\mathbb{R}\) control the relative weighting of the different objective components.
%The trade-off between feasibility and solution quality is controlled by the scaling parameter $\kappa_1$, while the two terms in the hybrid objective are weighted using a second scaling factor $\kappa_2$.
The asterisk indicates that the shooting time \(\tau_s\) and the final coast time \(\tau_f\) have been selected through a preliminary screening algorithm so as to minimize the objective.
The~\emph{\nameref{sec: objective function}} section describes this procedure in more detail.

\begin{figure}[b!]
\centering\includegraphics[width=0.8\textwidth]{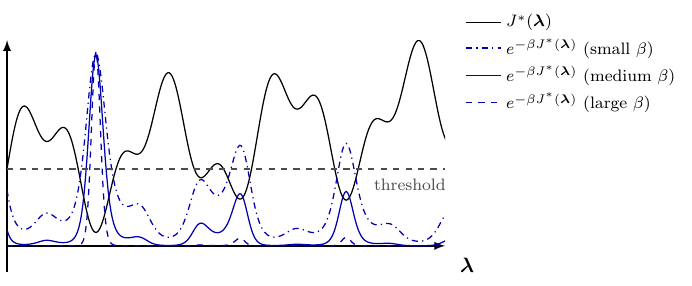}
	\caption{Reformulation of optimization as sampling: minima of \(J^*(\boldsymbol{\lambda})\) correspond to peaks of the unnormalized target density \(\pi(\boldsymbol{\lambda})\equiv\exp(-\beta J^*(\boldsymbol{\lambda}))\). The scaling factor $\beta$ controls the width and sharpness of the peaks.}
	\label{fig: sampling function}
\end{figure}

After the time variables have been fixed by the preliminary screening algorithm, the problem is an unconstrained minimization of the  objective \(J^*(\boldsymbol{\lambda};\alpha)\) over \(\boldsymbol{\lambda}\in\mathbb{R}^d\), where \(d=4\) for planar transfers and \(d=6\) for spatial transfers.
Equivalently, this optimization problem can be recast as a sampling problem over the same domain by defining the unnormalized target density
\begin{equation}
\label{eq: target distribution}
\pi(\boldsymbol{\lambda}\,\vert\,\alpha)\equiv \exp\!\left(-\beta J^*(\boldsymbol{\lambda};\alpha)\right),
\end{equation}
where the scaling factor \(\beta\) controls the width of the modes of the distribution and the steepness of its gradients.
Figure~\ref{fig: sampling function} illustrates this reformulation for a one-dimensional example and highlights the effect of \(\beta\).

Direct sampling from \(\pi(\boldsymbol{\lambda}\,\vert\,\alpha)\) for a given value of \(\alpha\) would require evaluation of the normalization constant \(\int_{\mathbb{R}^d} \pi(\boldsymbol{\lambda}\,\vert\,\alpha)\, d\boldsymbol{\lambda}\), which is intractable in practice.
Two common approaches for sampling from such distributions are MCMC methods and generative models.
The proposed framework combines diffusion models with several MCMC variants to leverage the complementary strengths of both approaches.
Diffusion models learn an approximation of the target distribution, conditioned on the mission parameter, from a finite set of training samples.
Their key advantage is that they learn a reusable representation of the distribution that can be sampled repeatedly at low cost.
However, this requires large amounts of training data across the parameter range of interest.
By contrast, MCMC methods construct a Markov chain that, under standard regularity conditions, converges asymptotically to the target distribution.
They therefore provide a practical mechanism for generating finite sets of samples from the target distribution, which can then be used as training data for the diffusion model.

In the general MCMC problem setting visualized in the right panel of Figure~\ref{fig: transfer_learning_graphic}, we assume that a baseline diffusion model has been trained to generate costate samples from \(\pi(\boldsymbol{\lambda}\,\vert\, \alpha=\alpha_1)\), and we seek to generate training samples for a new parameter value \(\alpha_2\).
Under the assumption that the solution distribution varies continuously with the mission parameter, samples from \(\pi(\boldsymbol{\lambda}\,\vert\, \alpha=\alpha_1)\) provide effective initial states for MCMC chains targeting \(\pi(\boldsymbol{\lambda}\,\vert\, \alpha=\alpha_2)\), provided that the homotopy step between \(\alpha_1\) and \(\alpha_2\) is chosen sufficiently small.
In this way, information learned at \(\alpha_1\) is transferred to the nearby target parameter \(\alpha_2\) through MCMC-based sample generation rather than by solving the new problem entirely from scratch.
To this end, \(N_\mathrm{chains}\) costates are first sampled from the baseline model and used to initialize \(N_\mathrm{chains}\) independent Markov chains.
Each chain is run for \(M\) iterations, with the first \(M_0\) samples discarded as burn-in, where \(M\) and \(M_0\) are chosen based on the complexity of the target distribution and the observed mixing behavior of the chains. 
The retained samples are treated as approximate draws from the target distribution and provide high-quality initial costates for the target parameter value. 

While these samples could in principle be used immediately to fine-tune the baseline model at \(\alpha_2\), the more efficient strategy adopted here is to repeat this procedure across a sequence of parameter values within a homotopy scheme. 
At each homotopy step, the final state of each chain from the previous parameter value is used to initialize the corresponding chain for the next one, while \(\alpha\) is advanced in fixed increments. 
The specific MCMC variants considered in this work are introduced in the \emph{\nameref{sec: gradient-based MCMC}} section.

After the desired number of homotopy steps, the resulting costate-reward pairs across all parameter values are used to fine-tune the baseline diffusion model through reward-weighted likelihood optimization.
This procedure is described in more detail in the \emph{\nameref{sec: fine-tuning}} section.
During fine-tuning, the diffusion model learns a representation of the target distribution conditioned on $\alpha$, enabling subsequent sampling for arbitrary values of $\alpha$ within the analyzed range.
These samples then provide high-quality initial guesses for a gradient-based solver, enabling rapid convergence to a local minimum.

The following subsections describe the individual components of the framework in more detail. 
We begin by outlining how diffusion models are used to learn a distribution over initial costates in the \emph{\nameref{sec: diffusion models}} section and how they are fine-tuned in the \emph{\nameref{sec: fine-tuning}} section. 
In the \emph{\nameref{sec: gradient-based MCMC}} section, we then present the MCMC sampling stage and the different MCMC variants considered in this work. 
Because these sampling methods rely on repeated evaluations of the objective function, or equivalently the target distribution, and, for gradient-based methods, its derivatives, we subsequently describe the \emph{\nameref{sec: objective function}} with a preliminary screening procedure, and the employed \emph{\nameref{sec: gradient approximation}}.

\subsection{Diffusion Models for Costate Generation}
\label{sec: diffusion models}

As part of the family of latent variable probabilistic models, diffusion models are capable of modeling complex, high-dimensional distributions.
In the proposed framework, diffusion models are used to learn a conditional distribution over high-quality initial costates \(\boldsymbol{\lambda}\) for LT transfers. 
Given a dataset of costate samples
$
\mathcal{D}=\{(\boldsymbol{\lambda}^{(i)}_0,\alpha^{(i)})\}_{i=1}^{N_\mathrm{data}}$ with
$\boldsymbol{\lambda}^{(i)}_0 \sim p_0(\boldsymbol{\lambda}\,\vert\,\alpha^{(i)}),
$
where the subscript $0$ denotes the initial state of the diffusion chain, the goal is to learn a representation of the underlying data distribution \(p_0(\boldsymbol{\lambda}\,\vert\,\alpha)\).
Assuming a high-quality dataset, \(p_0(\boldsymbol{\lambda}\,\vert\,\alpha)\) is close to the target distribution $\pi(\boldsymbol{\lambda}\,\vert\,\alpha)$.
The diffusion model can therefore be used to generate new costate samples that approximate draws from \(\pi(\boldsymbol{\lambda}\,\vert\,\alpha)\).
It thus provides a fast mechanism for generating diverse initial guesses for an indirect trajectory-optimization solver.

We employ a denoising diffusion probabilistic model (DDPM) following Ho et al.~\cite{Ho.6192020}. 
The model is defined through a forward diffusion process that gradually corrupts costate samples with Gaussian noise and a learned reverse process that reconstructs the costate distribution from noise. 
Starting from a costate sample \(\boldsymbol{\lambda}_0 \sim p_0(\boldsymbol{\lambda}\,\vert\,\alpha)\) and a prescribed number of diffusion steps $N$, the forward process generates increasingly noisy variables \(\boldsymbol{\lambda}_1,\dots,\boldsymbol{\lambda}_N\) according to
\begin{equation}
    q(\boldsymbol{\lambda}_n \,\vert\, \boldsymbol{\lambda}_{n-1})
    =
    \mathcal{N}\!\left(
    \boldsymbol{\lambda}_n;
    \sqrt{1-\beta_n}\,\boldsymbol{\lambda}_{n-1},
    \beta_n \boldsymbol{I}
    \right),
\end{equation}
where \(\{\beta_n\}_{n=1}^N\) is a prescribed variance schedule.
This implies the closed-form relation
$\boldsymbol{\lambda}_n = \sqrt{\overline{\alpha}_n} \boldsymbol{\lambda}_0 + \sqrt{1 - \overline{\alpha}_n} \boldsymbol{\epsilon}$, where $\boldsymbol{\epsilon}\sim\mathcal{N}(\boldsymbol{0},\boldsymbol{I})$, \(\alpha_n = 1-\beta_n\), and \(\overline{\alpha}_n = \prod_{i=1}^{n}\alpha_i\).
For sufficiently large \(N\), the terminal distribution approaches a standard normal distribution.

A neural network \(\boldsymbol{\epsilon}_{\theta}(\boldsymbol{\lambda}_n,n,\alpha)\), parameterized by \(\theta\), is trained to predict the Gaussian noise $\boldsymbol{\epsilon}$ added at each diffusion step.
The training objective is the standard DDPM loss
\begin{equation}
\label{eq: ddpm_loss_costate}
L(\theta)
=
\mathbb{E}_P
\left[
\left\|
\boldsymbol{\epsilon}-
\boldsymbol{\epsilon}_{\theta}(\boldsymbol{\lambda}_n,n,\alpha)
\right\|^2
\right],
\end{equation}
where $((\boldsymbol{\lambda}_0,\alpha), \boldsymbol{\epsilon}, n) \sim P \equiv \mathcal{D}\times\mathcal{N}(0, \boldsymbol{I})\times \text{Uniform}(\{1, \dots, N\})$. Thus, $(\boldsymbol{\lambda}_0,\alpha)$ is drawn from the training dataset $\mathcal{D}$, \(\boldsymbol{\epsilon}\) is Gaussian noise, and \(n\) is drawn uniformly from the diffusion steps.
In this way, the network learns to reverse the noising process and reconstruct costate samples conditioned on the mission parameter $\alpha$.

Due to its simplicity, classifier-free guidance~\cite{Ho.2022} (CFG) is used for conditional sample generation.
During training, $\alpha$ is replaced by a null token with a fixed probability so that the same network learns both conditional and unconditional denoising predictions.
At inference time, these two predictions are combined: 
\begin{equation}
\label{eq: cfg_ddpm_costate}
\boldsymbol{\epsilon}_{\theta}^{\mathrm{CFG}}(\boldsymbol{\lambda}_n,n,\alpha)
=
(1+w)\,\boldsymbol{\epsilon}_{\theta}(\boldsymbol{\lambda}_n,n,\alpha)
-
w\,\boldsymbol{\epsilon}_{\theta}(\boldsymbol{\lambda}_n,n,\varnothing),
\end{equation}
where $w$ is the guidance weight.
\begin{figure}[t!]
\centering
\includegraphics[width=.8\textwidth]{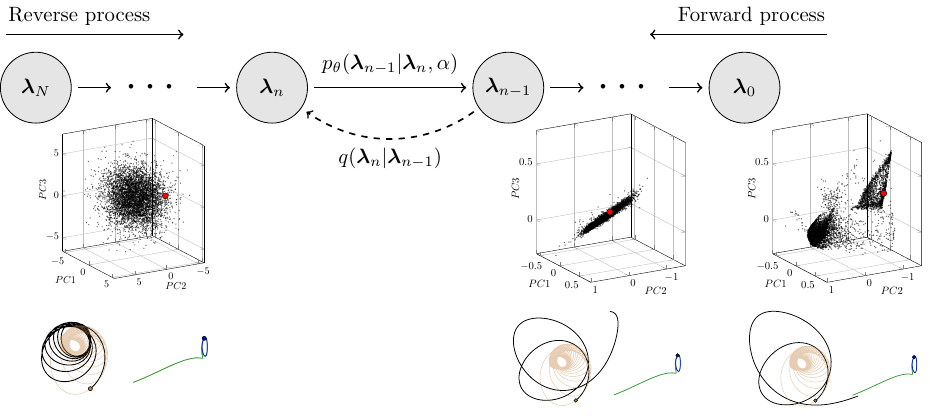}
\caption{Visualization of the forward and reverse diffusion processes for an indirect spacecraft trajectory optimization problem in the Earth-Moon system. 
The distributions depict costates in the space of their first three principal components ($PC1,PC2$ and $PC3$) at various stages of the diffusion process. 
An example trajectory corresponding to the realization of the data-point marked in red is shown below. Adapted from Ref.~\cite{graebner_JAS}.}
\label{fig: diffusion_process}
\end{figure}
During sampling, the reverse diffusion process is initialized from \(\boldsymbol{\lambda}_N \sim \mathcal{N}(\boldsymbol{0},\boldsymbol{I})\), and one step of the learned reverse transition \(p_{\theta}(\boldsymbol{\lambda}_{n-1}\,\vert\,\boldsymbol{\lambda}_n,\alpha)\) is given by
\begin{equation}
\label{eq: sampling}
\boldsymbol{\lambda}_{n-1} = \frac{1}{\sqrt{\alpha_n}}
\left(\boldsymbol{\lambda}_n-\frac{1-\alpha_n}{\sqrt{1-\overline{\alpha}_n}}\boldsymbol{\epsilon}_{\theta}^{\mathrm{CFG}}(\boldsymbol{\lambda}_n,n,\alpha)\right)+\sqrt{\tilde{\beta}_n}\,\boldsymbol{\xi}_n,
\end{equation}
where \(\boldsymbol{\xi}_n \sim \mathcal{N}(\boldsymbol{0},\boldsymbol{I})\) for \(n>1\) and \(\boldsymbol{\xi}_1=\boldsymbol{0}\).
Repeated sampling therefore provides a computationally efficient way to generate many diverse candidate initial costates for a prescribed mission parameter.
Figure~\ref{fig: diffusion_process} visualizes the forward and reverse processes for a spacecraft trajectory optimization example, where the data distribution is defined over the costate space of high-quality solutions.

\subsection{Reward-Weighted Fine-Tuning}
\label{sec: fine-tuning}
After the homotopy-MCMC stage has generated costate samples for new mission-parameter values, the baseline diffusion model is adapted to this data through supervised fine-tuning. 
The objective of this step is to improve the model's representation of the conditional costate distribution in the newly explored parameter regime.

The model parameters are initialized from the baseline diffusion model and updated using the newly generated dataset \(\mathcal{D}^{\mathrm{new}}\). 
Optionally, the original dataset \(\mathcal{D}\) can also be included as part of \(\mathcal{D}^{\mathrm{new}}\). %, either by explicitly computing reward values for those samples or by assigning them a common average weight.
Rather than weighting all samples equally, the fine-tuning procedure assigns greater importance to higher-quality costates through reward-weighted likelihood optimization~\cite{Lee.2232023}.
Using reward function \(R(\boldsymbol{\lambda};\alpha)\), the loss is 
\begin{equation}
\label{eq:updated loss}
L(\theta)
=
\mathbb{E}_{P^{\mathrm{new}}}
\Big[
R(\boldsymbol{\lambda}_0;\alpha)\,
\big\|
\boldsymbol{\epsilon}-\boldsymbol{\epsilon}_{\theta}(\boldsymbol{\lambda}_n,n,\alpha)
\big\|^2
\Big],
\end{equation}
where \(P^{\mathrm{new}}\) denotes the joint training distribution induced by the new dataset \(\mathcal{D}^{\mathrm{new}}\), analogous to the distribution \(P\) used for baseline diffusion-model training in Eq.~\eqref{eq: ddpm_loss_costate}. 
The reward is chosen as a rescaled version of the target distribution,
\begin{equation}
    \label{eq: reward}
    R(\boldsymbol{\lambda};\alpha)
    \equiv
    c_1\exp\!\left(-c_2 J^*(\boldsymbol{\lambda};\alpha)\right),
\end{equation}
where the problem-specific constants \(c_1\) and \(c_2\) are selected such that \(R(\boldsymbol{\lambda};\alpha) \in [0.1,1]\). 
As a result, costates with lower objective values contribute more strongly during fine-tuning. 
Since these reward values are directly available from the final MCMC step for each training sample, this weighting can be incorporated with negligible additional cost.

The resulting fine-tuned model provides an updated conditional generative representation that can be sampled efficiently across the expanded mission-parameter range.

%For both large language models and text-to-image systems, prior work has shown that fine-tuning with human feedback can significantly improve generative performance~\cite{Ouyang.342022,Lee.2232023,Ziegler.9182019}.
%In these methods, a separate reward model is first trained from human evaluations of generated samples. 
%This reward model is then used to score outputs from a pre-trained baseline model, and those scores guide the fine-tuning procedure through reward-weighted likelihood maximization~\cite{Lee.2232023}.
%

%In our setting, we adopt a similar strategy for fine-tuning diffusion models that generate initial costates for trajectory optimization.
%However, unlike the human-feedback paradigm, the reward function $R(\boldsymbol{z})$ is directly computable, (see \eqref{eq: reward}), eliminating the need to train a separate reward model.

\subsection{MCMC Methods for Costate Sampling}
\label{sec: gradient-based MCMC}

MCMC methods generate samples from a target distribution with unnormalized density \(\pi(\boldsymbol{\lambda})\) by constructing a Markov chain whose stationary distribution is \(\pi(\boldsymbol{\lambda})\). 
In the proposed framework, the target distribution is a conditional distribution, and these methods are used to generate costate samples for fixed values of the conditional parameter $\alpha$. 
The three algorithms considered here all belong to the class of Metropolis-Hastings methods.
At each iteration $k$, with current costate $\boldsymbol{\lambda}^{k}$, a candidate costate $\tilde{\boldsymbol{\lambda}}^{k+1}$ is drawn from a proposal distribution $q(\tilde{\boldsymbol{\lambda}}^{k+1};\boldsymbol{\lambda}^{k})$, and this proposal is then accepted or rejected according to an acceptance probability $a(\tilde{\boldsymbol{\lambda}}^{k+1},\boldsymbol{\lambda}^{k})$. 
With a suitable choice of the acceptance probability, the transition kernel preserves the target distribution, and under standard regularity conditions, the chain converges asymptotically to $\pi(\boldsymbol{\lambda})$.
Standard methods such as RWM explore the state space using only local random perturbations and can therefore mix slowly, especially when the target distribution is multimodal or strongly anisotropic~\cite{Roberts.1998}. %, Tran.4202025}. 
Gradient-based samplers address this limitation by incorporating \(\nabla \log \pi(\boldsymbol{\lambda})\) into the proposal mechanism, allowing the chain to better exploit local information about the geometry of the target distribution.

\subsubsection{Random-Walk Metropolis}
The RWM algorithm~\cite{Metropolis.1953,Hastings.1970} serves as the baseline MCMC method in this work.
Starting from the current state \(\boldsymbol{\lambda}^k\), a proposal is generated by drawing a perturbation from a symmetric distribution
\begin{equation}
    \label{eq: proposal distribution}
    q(\tilde{\boldsymbol{\lambda}}^{k+1}; \boldsymbol{\lambda}^k) 
    \equiv \mathcal{N}(\tilde{\boldsymbol{\lambda}}^{k+1}; \boldsymbol{\lambda}^k,\boldsymbol{\Sigma}).
\end{equation}
In Eq.~\eqref{eq: proposal distribution}, \(\boldsymbol{\Sigma}\) denotes the proposal covariance, which should ideally be approximately proportional to the covariance of the target distribution. 
Due to the symmetry of the proposal distribution, the Metropolis-Hastings acceptance probability reduces to
\begin{equation}
    a(\tilde{\boldsymbol{\lambda}}^{k+1},\boldsymbol{\lambda}^k)
    =
    \frac{\pi(\tilde{\boldsymbol{\lambda}}^{k+1})}{\pi(\boldsymbol{\lambda}^k)}
    \wedge 1,
\end{equation}
where \(a\wedge b\) denotes \(\min\{a,b\}\).
%If the proposal is accepted, the chain moves to \(\tilde{\boldsymbol{\lambda}}^{k+1}\); otherwise it remains at \(\boldsymbol{\lambda}^k\).
%Although RWM is simple to implement and requires only evaluations of the target density, its exploration can be inefficient in high-dimensional or highly structured target distributions, which motivates the use of gradient-based proposals.

\subsubsection{Metropolis-Adjusted Langevin Algorithm}
MALA proposes new states using Langevin dynamics and accepts them based on the Metropolis-Hastings algorithm~\cite{Roberts.1998, Metropolis.1953, Hastings.1970}.
%Langevin dynamics are described by the stochastic differential equation
%\begin{equation}
%d\boldsymbol{\lambda}_t
%= \nabla \log \pi(\boldsymbol{\lambda}_t)\, dt
%\;+\; \sqrt{2}\, d\boldsymbol{w}_t ,
%\end{equation}
%where $\boldsymbol{w}_t$ is standard Brownian motion.
%As $t\rightarrow\infty$, the probability distribution of $\boldsymbol{\lambda}(t)$ converges to the stationarity distribution $\pi$, which is invariant under the diffusion~\cite{Roberts.1996}.
A first-order Euler-Maruyama discretization of these stochastic dynamics yields the standard MALA proposal distribution
\begin{equation}
    \label{eq: proposal distribution basic}
    q(\tilde{\boldsymbol{\lambda}}^{k+1}; \boldsymbol{\lambda}^k) 
    \equiv \mathcal{N}(\tilde{\boldsymbol{\lambda}}^{k+1}; \boldsymbol{\lambda}^k+\dfrac{\epsilon}{2}\nabla\!\log(\pi(\boldsymbol{\lambda}^{k})),\epsilon\boldsymbol{I}),
\end{equation}
with timestep $\epsilon$ and identity matrix $\boldsymbol{I}$.
This basic MALA proposal performs poorly if the parameter space exhibits anisotropic scaling.
We therefore use the modified proposal 
\begin{equation}
    \label{eq: proposal distribution MALA}
    q(\tilde{\boldsymbol{\lambda}}^{k+1}; \boldsymbol{\lambda}^k)
    \equiv 
    \mathcal{N}\!\left(
        \tilde{\boldsymbol{\lambda}}^{k+1};
        \boldsymbol{\lambda}^k
        + \frac{\epsilon}{2}\,\boldsymbol{\Sigma}^{1/2}
        \frac{\nabla \log\!\pi(\boldsymbol{\lambda}^k)}
        {\left\lVert \nabla \log\!\pi(\boldsymbol{\lambda}^k) \right\rVert_2},
        \;\boldsymbol{\Sigma}
    \right),
\end{equation}
where, similar to RWM, the proposal covariance $\boldsymbol{\Sigma}$ is ideally approximately proportional to the covariance of the target distribution. 
Based on empirical results, we normalize the gradient to prevent excessively large drift steps in regions where the objective landscape is very steep.
For $\epsilon=0$ this proposal reduces exactly to RWM.
For $\epsilon \neq0$ the method augments the random walk with a gradient drift step before drawing the stochastic proposal.  
The modified proposal allows us to control the size of the gradient step and the random step independently, which is not possible for the standard proposal in Eq.~\eqref{eq: proposal distribution basic}.
A proposed state is accepted with probability
\begin{equation}
    a(\tilde{\boldsymbol{\lambda}}^{k+1},\boldsymbol{\lambda}^k)=\frac{\pi(\tilde{\boldsymbol{\lambda}}^{k+1})q\bigl(\boldsymbol{\lambda}^{k};\tilde{\boldsymbol{\lambda}}^{k+1}\bigr)}{\pi(\boldsymbol{\lambda}^{k})q\bigl(\tilde{\boldsymbol{\lambda}}^{k+1};\boldsymbol{\lambda}^{k}\bigr)}\wedge 1.
\end{equation}
%The step variance is an important parameter with a larger value accelerating the mixing process but decreasing acceptance probabilities.
%Roberts and Rosenthal~\cite{Roberts.1998} have shown that the asymptotically optimal acceptance rate is given by $0.574$, with values in between $0.4$ and $0.8$ generally yielding good results. 

\subsubsection{Hamiltonian Monte Carlo}
HMC is a gradient-based method that uses Hamiltonian dynamics to construct an MCMC algorithm. 
By computing new states along trajectories under these dynamics, the correlation between successive samples is reduced and exploration of the state space is accelerated.
Because the trajectories follow regions of high target density, proposed states are accepted with high probability~\cite{Neal.2012}.
We introduce the potential energy of the target distribution as $U(\boldsymbol{\lambda})=-\log \pi(\boldsymbol{\lambda})$.
By augmenting the state with additional momentum variables $\boldsymbol{p}\in\mathbb{R}^d$, where $d$ is the dimension of $\boldsymbol{\lambda}$, and introducing a kinetic energy term $K(\boldsymbol{p})$, the Hamiltonian of the system is defined as 
\begin{equation}
    H(\boldsymbol{\lambda},\boldsymbol{p})=U(\boldsymbol{\lambda})+K(\boldsymbol{p}).
\end{equation}
A common choice is a homogeneous quadratic kinetic energy $K(\boldsymbol{p})=\frac{1}{2}\boldsymbol{p}^\top \boldsymbol{M}^{-1} \boldsymbol{p}$, with the mass matrix $\boldsymbol{M}$ typically taken to be diagonal~\cite{Neal.2012}.
The specific choice of the diagonal elements $\boldsymbol{M}_{ii}$ is an important degree of freedom in HMC
In practice, a good approach is to set $\boldsymbol{M}^{-1}$ proportional to the covariance of the target distribution or to an estimate thereof~\cite{betancourt2018conceptualintroductionhamiltonianmonte}. 
The dynamics of the augmented state are described by Hamilton's equations
\begin{equation}
\frac{\mathrm{d}\boldsymbol{\lambda}}{\mathrm{d}t}
  = \frac{\partial H}{\partial \boldsymbol{p}}
  = \nabla_{\boldsymbol{p}} K=\boldsymbol{M}^{-1} \boldsymbol{p},
\qquad
\frac{\mathrm{d}\boldsymbol{p}}{\mathrm{d}t}
  = -\,\frac{\partial H}{\partial \boldsymbol{\lambda}}
  = \nabla_{\boldsymbol{\lambda}}\log \pi(\boldsymbol{\lambda}).
\end{equation}

The first step of HMC is to sample the initial momentum at the current state from the normal distribution $\boldsymbol{p}^k\sim\mathcal{N}(\boldsymbol{p};\boldsymbol{0},\boldsymbol{M})$.
In the second step, a new state $(\tilde{\boldsymbol{\lambda}}^{k+1},\tilde{\boldsymbol{p}}^{k+1})$ is proposed by integrating the current state $(\boldsymbol{\lambda}^{k},\boldsymbol{p}^{k})$ under the dynamics defined by the Hamiltonian.
Numerically, this is carried out using a volume-preserving symplectic integrator such as the leapfrog method~\cite{Neal.2012}. 
As in the MALA method, this step requires the gradient $\nabla \log\pi(\boldsymbol{\lambda})$.
One iteration of the leapfrog method consists of the following steps:
\begin{align}
\boldsymbol{p}^k\!\left(t + \frac{\Delta t}{2}\right)
&= \boldsymbol{p}^k(t)
  + \frac{\Delta t}{2}\,\nabla \log\pi(\boldsymbol{\lambda})
    \big|_{\boldsymbol{\lambda}=\boldsymbol{\lambda}^k(t)},
\label{eq:hmc_step1} \\
\boldsymbol{\lambda}^k(t+\Delta t)
&= \boldsymbol{\lambda}^k(t)
  + \Delta t\, \boldsymbol{M}^{-1}\,
    \boldsymbol{p}^k\!\left(t + \frac{\Delta t}{2}\right),
\label{eq:hmc_step2} \\
\boldsymbol{p}^k(t+\Delta t)
&= \boldsymbol{p}^k\!\left(t + \frac{\Delta t}{2}\right)
  + \frac{\Delta t}{2}\,\nabla \log\pi(\boldsymbol{\lambda})
    \big|_{\boldsymbol{\lambda}=\boldsymbol{\lambda}^k(t+\Delta t)},
\label{eq:hmc_step3}
\end{align}
with initial samples $\boldsymbol{p}^k(0)=\boldsymbol{p}^k$ and $\boldsymbol{\lambda}^k(0)=\boldsymbol{\lambda}^k$.
The integration runs for a total time of $L\Delta t$, where $L$ is the number of integration steps and $\Delta t$ is the integration timestep. 
The proposed state for the Metropolis-Hastings step is given by $\tilde{\boldsymbol{\lambda}}^{k+1}=\boldsymbol{\lambda}^{k}(L\Delta t)$ and $\tilde{\boldsymbol{p}}^{k+1}=\boldsymbol{p}^{k}(L\Delta t)$.
The acceptance probability of this new state is then
\begin{equation}
    a(\tilde{\boldsymbol{\lambda}}^{k+1},\boldsymbol{\lambda}^k)=\frac{\exp{(-H(\tilde{\boldsymbol{\lambda}}^{k+1},\tilde{\boldsymbol{p}}^{k+1}))}}{\exp(-{H(\boldsymbol{\lambda}^{k},\boldsymbol{p}^{k}))}}\wedge1.
\end{equation}
Based on empirical observations, we adapt the standard HMC update to use the same normalized gradient scaling employed in the MALA method. 
Specifically, in Eqs.~\eqref{eq:hmc_step1} and~\eqref{eq:hmc_step3}, we replace $\nabla \log\pi(\boldsymbol{\lambda})$ by $\sqrt{\boldsymbol{M}}\frac{\nabla \log\pi(\boldsymbol{\lambda})}{\left\lVert \nabla \log\!\pi(\boldsymbol{\lambda}) \right\rVert_2}$.
With this modification, the HMC proposal becomes exactly equivalent to a MALA proposal when $L=1$, $\Delta t=\epsilon$, and $\boldsymbol{M}=\epsilon^2\boldsymbol{\Sigma}^{-1}$. 
To keep the methods comparable, we parameterize the HMC algorithm directly in terms of $\epsilon$ and $\boldsymbol{\Sigma}$, and then compute $\Delta t$ and $\boldsymbol{M}$ from these quantities.

\subsection{Objective Function Evaluation}
\label{sec: objective function}
Each MCMC iteration requires evaluating the target density $\pi(\boldsymbol{\lambda}\,\vert\,\alpha)$ (defned in Eq.~\eqref{eq: target distribution}) at a given costate sample, which reduces to evaluating the objective function \(J^*(\boldsymbol{\lambda};\alpha)\).
This subsection describes how \(J^*\) is computed through a preliminary screening procedure based on a modified version of the algorithm introduced in our previous work~\cite{graebner_JAS}.

Starting from the fixed initial state and a given initial costate \(\boldsymbol{\lambda}\), the trajectory is propagated under the system dynamics in Eq.~\eqref{Equation: Combined state equations} for a maximum shooting time \(\tau_{s,\max}\).
The propagated trajectory is parameterized by the shooting time \(\tau_s\), while the target point on the final orbit is parameterized by a coast time \(\tau_f\).
Here, \(\tau_f\) denotes the time required to reach a particular point on the target orbit and is therefore restricted by the orbit period \(\mathcal{T}_f\).
This allows the corresponding terminal position and velocity on the target orbit to be written as \(\boldsymbol{r}_f(\tau_f)\) and \(\boldsymbol{v}_f(\tau_f)\).

For fixed \(\boldsymbol{\lambda}\), \(\tau_s\), and \(\tau_f\), we define the screening objective
\begin{equation}
    \label{eq: general objective}
    J(\boldsymbol{\lambda},\tau_s,\tau_f;\alpha)
    =
    e(\boldsymbol{\lambda},\tau_s,\tau_f;\alpha)
    +
    \kappa_1\left(
        \frac{\Delta m(\boldsymbol{\lambda},\tau_s;\alpha)}{m_0}
        +
        \kappa_2 \tau_s
    \right),
\end{equation}
where
\begin{equation}
\label{eq: constr violation}
e(\boldsymbol{\lambda},\tau_s,\tau_f;\alpha)
=
\left\Vert
    \begin{pmatrix}
     \boldsymbol{r}_f(\tau_f) - \boldsymbol{r}(\tau_s) \\
     \boldsymbol{v}_f(\tau_f) - \boldsymbol{v}(\tau_s)
    \end{pmatrix}
\right\Vert_2.
\end{equation}
This screening objective evaluates the quality of a costate sample together with a particular choice of shooting and coast times.
For each costate sample \(\boldsymbol{\lambda}\), the screening algorithm selects the shooting and coast times that minimize this quantity:
\begin{equation}
    (\tau_s^*(\boldsymbol{\lambda};\alpha),\,\tau_f^*(\boldsymbol{\lambda};\alpha))
    =
    \argmin_{\substack{0 \le \tau_s \le \tau_{s,\max} \\[2pt] 0 \le \tau_f \le \mathcal{T}_f}}
    J(\boldsymbol{\lambda},\tau_s,\tau_f;\alpha).
\end{equation}
Substituting these minimizing values into Eq.~\eqref{eq: general objective} yields the objective function that was already introduced in Eq.~\eqref{eq: objective} and is used throughout the remainder of the framework:
\begin{equation}
    \label{eq:objective function}
    J^*(\boldsymbol{\lambda};\alpha)
    =
    J(\boldsymbol{\lambda},\tau_s^*(\boldsymbol{\lambda};\alpha),\tau_f^*(\boldsymbol{\lambda};\alpha);\alpha)
    =
    \min_{\substack{0\le \tau_s\le \tau_{s,\max} \\ 0\le \tau_f\le \mathcal{T}_f}}
    J(\boldsymbol{\lambda},\tau_s,\tau_f;\alpha).
\end{equation}

Within the algorithm, this minimization is carried out as a nearest-neighbor search over discretized sets of states on the propagated trajectory and the final orbit.
It is implemented efficiently using a $k$-dimensional ($k$-d) tree search~\cite{Bentley.1975}.
Accordingly, the MCMC methods operate on the objective function \(J^*(\boldsymbol{\lambda};\alpha)\), while the dependence on \(\tau_s\) and \(\tau_f\) is handled implicitly through the preliminary screening step.

\subsection{Gradient Approximation}
\label{sec: gradient approximation}
Each iteration of the gradient-based MCMC algorithms also requires evaluation of
$\nabla \log \pi(\boldsymbol{\lambda})=-\beta \nabla J^*(\boldsymbol{\lambda})$,
which reduces to computing \(\nabla J^*(\boldsymbol{\lambda}^k)\) at the current costate sample \(\boldsymbol{\lambda}^k\).
For brevity, we omit the dependence on \(\alpha\) throughout this section.
Although the objective function \(J^*\) is continuous, it is not continuously differentiable, and its gradient need not exist everywhere.
To obtain a differentiable surrogate, we fix the shooting and coast times at the current sample $
\tau_s^k=\tau_s^*(\boldsymbol{\lambda}^k)$, $\tau_f^k=\tau_f^*(\boldsymbol{\lambda}^k),$
and define the frozen-time objective
\begin{equation}
    J^k(\boldsymbol{\lambda}) = J(\boldsymbol{\lambda},\tau_s^k,\tau_f^k),
\end{equation}
which is continuously differentiable with respect to \(\boldsymbol{\lambda}\).

\begin{figure}[b!]
\centering\includegraphics[width=0.8\textwidth]{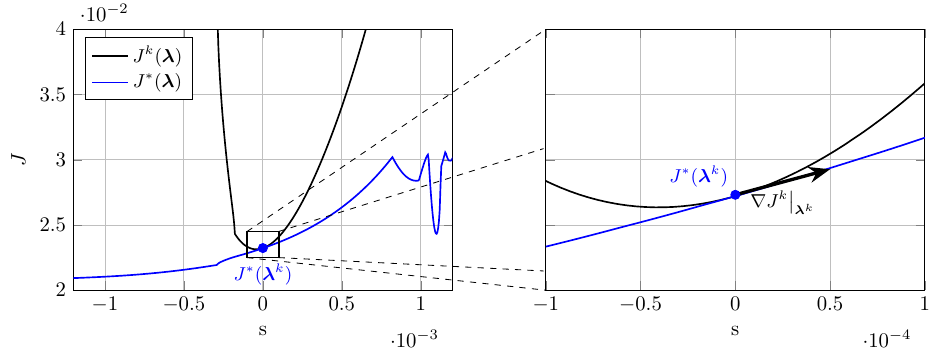}
\caption{\(J^k\) and \(J^*\) for a randomly selected sample \(\boldsymbol{\lambda}^k\). The objective values are plotted along the line \(\boldsymbol{\lambda}^k + s\,\nabla J^k(\boldsymbol{\lambda})|_{\boldsymbol{\lambda}=\boldsymbol{\lambda}^k}\), where \(s\in\mathbb{R}\). The right panel shows a close-up near \(s=0\).}
\label{fig:Grad_approx}
\end{figure}

We now show that $\boldsymbol{d}=-\nabla J^k(\boldsymbol{\lambda})\big|_{\boldsymbol{\lambda}=\boldsymbol{\lambda}^k}$
is a descent direction for \(J^*\) at \(\boldsymbol{\lambda}^k\) whenever \(\boldsymbol{d}\neq \boldsymbol{0}\).
First, by definition,
\begin{equation}
    \label{eq: step 1}
    J^*(\boldsymbol{\lambda}^k)=J^k(\boldsymbol{\lambda}^k),
\end{equation}
and
\begin{equation}
    \label{eq: step 2}
    J^*(\boldsymbol{\lambda})\leq J^k(\boldsymbol{\lambda})
    \qquad \forall\,\boldsymbol{\lambda}.
\end{equation}
Since \(\boldsymbol{d}\) is the negative gradient of \(J^k\) at \(\boldsymbol{\lambda}^k\), it is a descent direction for \(J^k\), so there exists \(\overline{\eta}>0\) such that
\begin{equation}
\label{eq: step 3}
    J^k(\boldsymbol{\lambda}^k+\eta \boldsymbol{d})
    <
    J^k(\boldsymbol{\lambda}^k)
    \qquad \forall\,\eta\in(0,\overline{\eta}).
\end{equation}
Combining Eqs.~\eqref{eq: step 1}-\eqref{eq: step 3} yields
\begin{equation}
    J^*(\boldsymbol{\lambda}^k+\eta \boldsymbol{d})
    \leq
    J^k(\boldsymbol{\lambda}^k+\eta \boldsymbol{d})
    <
    J^k(\boldsymbol{\lambda}^k)
    =
    J^*(\boldsymbol{\lambda}^k)
    \qquad \forall\,\eta\in(0,\overline{\eta}),
\end{equation}
which establishes \(\boldsymbol{d}\) as a descent direction of \(J^*\) at \(\boldsymbol{\lambda}^k\).

Figure~\ref{fig:Grad_approx} compares \(J^*\) and \(J^k\) for a representative sample \(\boldsymbol{\lambda}^k\) along the direction of \(\nabla J^k\).
The zoomed view indicates that, at least locally for this sample, the frozen-time gradient provides a useful approximation for guiding descent in the true objective landscape.
Accordingly, during each MCMC step we approximate \(\nabla J^*(\boldsymbol{\lambda}^k)\) by \(\nabla J^k(\boldsymbol{\lambda})|_{\boldsymbol{\lambda}=\boldsymbol{\lambda}^k}\), which can be expressed directly in terms of the submatrices of the STM introduced in Eq.~\eqref{eq: STM G1 and G2}:
\begin{equation}
\begin{aligned}
    \nabla J^k(\boldsymbol{\lambda})\big|_{\boldsymbol{\lambda}=\boldsymbol{\lambda}^k}
    &=
    \nabla e(\boldsymbol{\lambda},\tau_s^k,\tau_f^k)\big|_{\boldsymbol{\lambda}=\boldsymbol{\lambda}^k}
    -\frac{\kappa_1}{m_0}\nabla m_f(\boldsymbol{\lambda},\tau_s^k)\big|_{\boldsymbol{\lambda}=\boldsymbol{\lambda}^k} \\
    &=
    \frac{1}{e(\boldsymbol{\lambda}^k,\tau_s^k,\tau_f^k)}
    \boldsymbol{G}_2^{\!\top}\boldsymbol{e}(\boldsymbol{\lambda}^k,\tau_s^k,\tau_f^k)
    +\frac{\kappa_1}{m_0}\boldsymbol{G}_1^{\!\top}.
\end{aligned}
\end{equation}
Using analytic derivatives obtained from the STM improves accuracy relative to numerical finite differencing, which is particularly important for multi-revolution transfers that are highly sensitive to small perturbations~\cite{Russell.2007}.
For a problem with \(d\) propagated states, inclusion of the STM increases the propagated state dimension by \(d^2\).

\begin{figure}[t!]
\centering\includegraphics[width=0.95\textwidth]{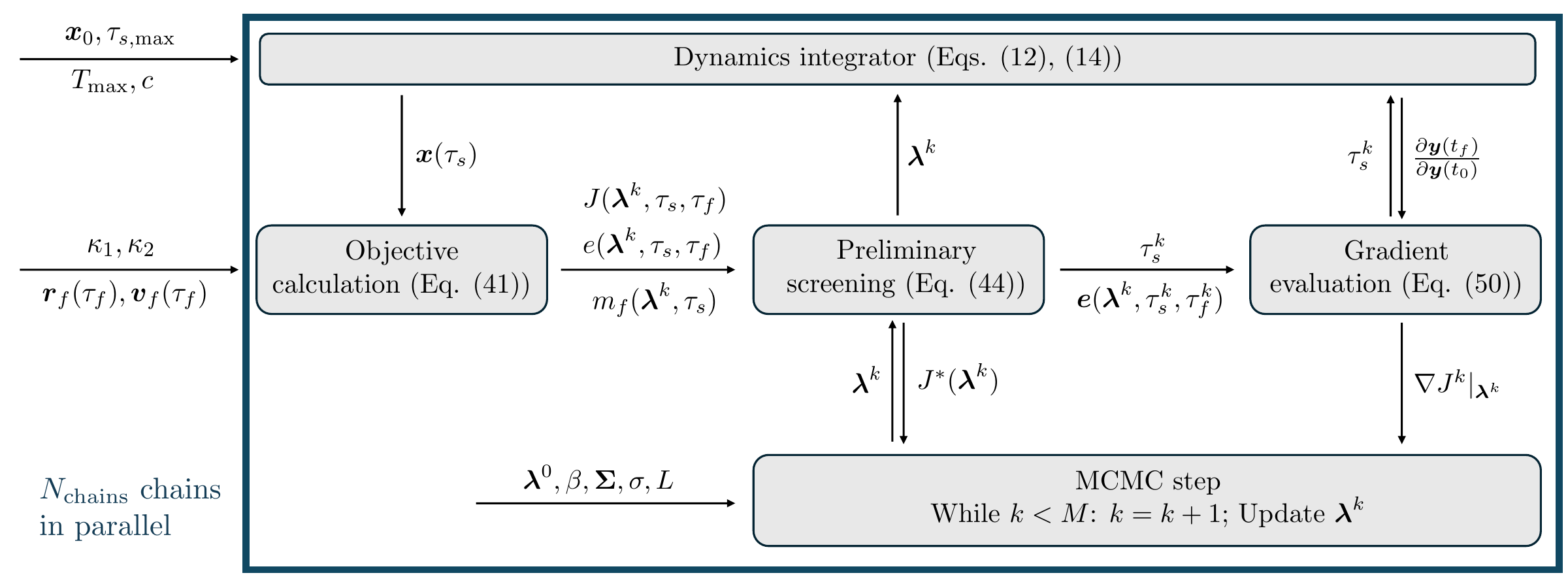}
	\caption{Flowchart of variables and equations for gradient-based MCMC sampling.}
	\label{fig: Mala workflow}
\end{figure}

Figure~\ref{fig: Mala workflow} summarizes the sequence of computations performed within one iteration of the gradient-based MCMC algorithms. 
Starting from the current costate sample, the trajectory is propagated, the screening objective is evaluated to determine the objective function $J^*$, and the corresponding frozen-time gradient is computed before carrying out the MALA or HMC update.

\section{Results}
\label{sec: results}
Results are presented for an LT transfer in the CR3BP, using the mass ratio $\mu$ as the mission parameter. 
After describing the test problem, we demonstrate why it is challenging and explain why MCMC is better suited to the proposed transfer-learning framework than a standard gradient-based solver. 
We then compare the performance of the three MCMC algorithms and benchmark the transfer-learning approach against generating solutions from scratch for the final $\mu$-value with a state-of-the-art method. 
Performance is evaluated in terms of feasibility ratio, objective quality, sample diversity, and computational cost. 
Finally, we visualize the distribution obtained after fine-tuning a diffusion model with samples generated by MALA.

Numerical integration of trajectories and STMs is performed using \texttt{pydylan}, the Python interface to the astrodynamics software package Dynamically Leveraged (N) Multibody Trajectory Optimization (DyLAN)~\cite{Beeson.Aug.2022}, which employs an adaptive-stepsize RK54 integrator.
We use a relative tolerance $10^{-12}$ and maximum step size $10^{-2}$ in NU for the adaptive integrator.
The switching function is interpolated to an accuracy of $10^{-13}$. 

\subsection{Problem Description}
\begin{table}[b!]
\centering
\caption{Problem parameters for Europa and Titan DRO transfers.}
\setlength{\tabcolsep}{6pt}
\begin{tabular}{lcc}
\toprule
\textbf{Trajectory parameters} & \textbf{Europa DRO} & \textbf{Titan DRO} \\
\midrule
Initial state $[\boldsymbol{r}_0^\top,\boldsymbol{v}_0^\top]$ [NU] 
    & $[1.0752, 0.0, 0.0, 0.0, -0.1499, 0.0]$
    & $[1.0758, 0.0, 0.0, 0.0, -0.1684, 0.0]$ \\
Terminal state $[\boldsymbol{r}_f^\top,\boldsymbol{v}_f^\top]$ [NU] 
    & $[1.0306, 0.0, 0.0, 0.0, -0.0727, 0.0]$
    & $[1.0304, 0.0, 0.0, 0.0, -0.1248, 0.0]$ \\
Orbital period target DRO $\mathcal{T}_f$ [TU] 
    & 4.1055 & 4.6558 \\
Max.\ shooting time $\tau_{s,\max}$ [TU] 
    & \multicolumn{2}{c}{90} \\
\midrule
\textbf{Spacecraft parameters} &  &  \\
\midrule
Initial mass $m_0$ [kg] 
    & \multicolumn{2}{c}{25,000} \\
Fuel mass [kg] 
    & \multicolumn{2}{c}{15,000} \\
Dry mass [kg] 
    & \multicolumn{2}{c}{10,000} \\
Specific impulse $I_\mathrm{sp}$ [s] 
    & 7,365 & 2,987 \\
Thrust magnitude $T_\mathrm{max}$ [N] 
    & 4.735 & 0.4500 \\
\midrule
\textbf{Natural units} & \textbf{Jupiter–Europa} & \textbf{Saturn–Titan} \\
\midrule
Distance unit [km] 
    & 670,900 & 1,221,870 \\
Time unit [s] 
    & 48,822.76 & 219,277.51 \\
Mass unit [kg] 
    & $1.898 \times 10^{27}$ & $5.685 \times 10^{26}$ \\
Mass parameter $\mu$ 
    & $2.528 \times 10^{-5}$ & $2.366 \times 10^{-4}$ \\
\bottomrule
\end{tabular}
\label{tab:comparison_Europa_Titan}
\end{table}
We test the framework on an LT transfer in the CR3BP and consider the mass parameter $\mu$ as the varying mission parameter $\alpha$. 
As the baseline model, we use a diffusion model from previous work~\cite{graebner_JAS} that was trained for a single value of \(\mu\), corresponding to the Jupiter-Europa system.
Through the transfer learning scheme we aim to generate data 
that spans the $\mu$-regime up to the Saturn-Titan system.
We therefore introduce $\alpha$ as a rescaled mass parameter:
\begin{equation}
\alpha=\frac{\mu-\mu_{\mathrm{JE}}}{\mu_{\mathrm{ST}}-\mu_{\mathrm{JE}}},
\quad \mu_{\mathrm{JE}}=2.525\times10^{-5},\;
\mu_{\mathrm{ST}}=2.366\times10^{-4},
\end{equation}
so that \(\alpha=0\) corresponds to Jupiter–Europa and \(\alpha=1\) to Saturn–Titan.
For intermediate values of \(\alpha\), we define artificial CR3BP systems by linearly interpolating the natural units between the two endpoint systems. 

The LT trajectory optimization problem is a planar multi-revolution transfer from a high DRO to a lower DRO around the secondary body.
The initial problem in the Jupiter-Europa system derives from the Jupiter Icy Moons Orbiter (JIMO) mission concept, which was canceled in 2005.
It was studied extensively by Russell~\cite{Russell.2007}, among others.
All relevant trajectory, spacecraft, and unit parameters for the two endpoint systems are summarized in Table~\ref{tab:comparison_Europa_Titan}. 
The spacecraft parameters are chosen to be identical in natural units for all systems, which leads to the different values in SI units presented in Table~\ref{tab:comparison_Europa_Titan}.
In all systems, the initial and target DROs are defined such that their crossings of the \(r_1\)-axis closest to the primary have the same distance from the secondary body in NU. 
The corresponding \(v_2\) velocities are then obtained by differential correction to produce closed orbits at each homotopy step.
\begin{figure}[t!]
  \centering
  %
  %--- Left: Europa -------------------------------------------
  \begin{subfigure}[b]{0.45\textwidth}  % width is free to vary
    \centering
    \includegraphics[height=5.9cm]{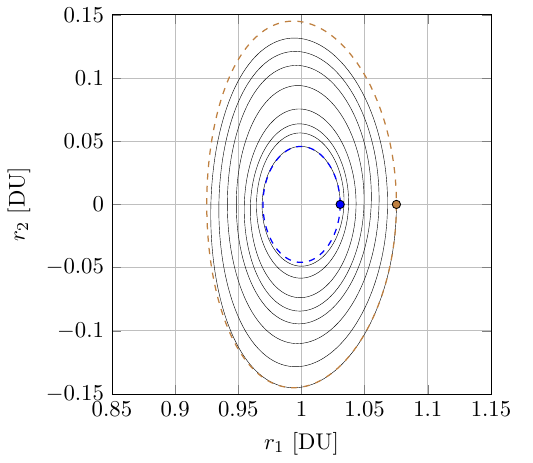}
    %\caption{Europa case}
    %\label{fig:traj_europa}
  \end{subfigure}
  \hfill
  %--- Right: Saturn ------------------------------------------
  \begin{subfigure}[b]{0.54\textwidth}  % can be narrower/wider as needed
    \centering
    \includegraphics[height=5.9cm]{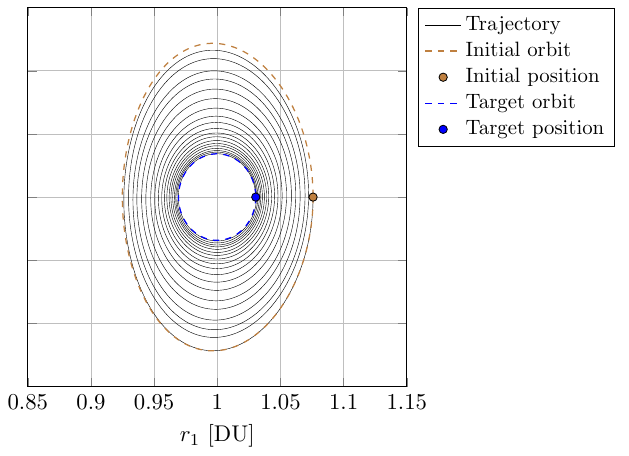}
    %\caption{Saturn case}
    %\label{fig:traj_saturn}
  \end{subfigure}

  \caption{Representative locally optimal trajectories for the Europa DRO transfer (left) and Titan DRO transfer (right).}
  \label{fig:traj_examples}
\end{figure}

Because the Saturn-Titan system has a mass parameter roughly one order of magnitude larger than that of Jupiter-Europa, the resulting DRO geometry differs substantially, as illustrated in Figure~\ref{fig:traj_examples}.
This difference also affects the structure of the transfer trajectories: solutions in the final Titan dataset complete an average of $23.7$ revolutions about the secondary body, compared to $10.3$ for Europa. 
The increase is driven by the larger energy gap between departure and arrival orbits, which is $4.11\times10^{-3}$ NU for Saturn-Titan versus $2.38\times10^{-3}$ NU for Jupiter-Europa. 
As a result, the Titan DRO transfer is more sensitive to small perturbations in the initial costates, producing a highly nonconvex objective landscape with many local minima, small basins of attraction, and steep gradients.

Although the system mass parameter would not typically vary significantly as a mission parameter during preliminary mission design, we use it here as a challenging demonstrative example. 
This setting is qualitatively similar to conditioning on the spacecraft maximum thrust magnitude, since in both cases the number of revolutions required to reach the target orbit changes substantially as the mission parameter varies. 
The present example is arguably more difficult, because increasing $\mu$ also increases the nonlinearity of the dynamics.

\subsection{Difficulty of the Problem and Motivation for MCMC}
Although the Europa and Titan DRO transfers may appear similar, generating solutions for the latter from those for the former is not a trivial task. 
This motivates both the homotopy and MCMC components of the proposed transfer-learning framework. 
A naive first attempt is to use existing Europa DRO solutions directly to initialize a gradient-based solver for the Titan DRO transfer. 
For this test, indirect trajectory optimization is performed using the \texttt{pydylan}~\cite{Beeson.Aug.2022} software, which applies a forward shooting transcription and solves the resulting nonlinear program using the sequential quadratic programming solver SNOPT~\cite{Gill.2005}.
Doing this for $1000$ initial solutions using analytic derivatives and a maximum solver time of $240\,\mathrm{s}$ in SNOPT yields no feasible solutions. 
This indicates that the solutions to the two problems are located in substantially different regions of costate space, making the transfer-learning task nontrivial.

\begin{figure}[t!]
  \centering

  \begin{subfigure}[b]{0.245\textwidth}
    \centering
    \includegraphics[height=3.65cm,keepaspectratio]{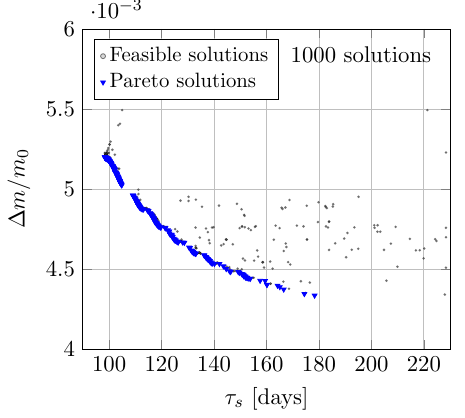}
    \caption{$\alpha=0.0$}
    \label{fig:dvtof_h0}
  \end{subfigure}\hfill
  \begin{subfigure}[b]{0.245\textwidth}
    \centering
    \includegraphics[height=3.65cm,keepaspectratio]{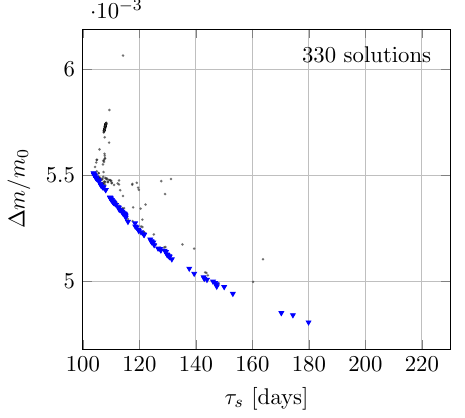}
    \caption{$\alpha=0.4$}
    \label{fig:dvtof_h04}
  \end{subfigure}\hfill
  \begin{subfigure}[b]{0.245\textwidth}
    \centering
    \includegraphics[height=3.65cm,keepaspectratio]{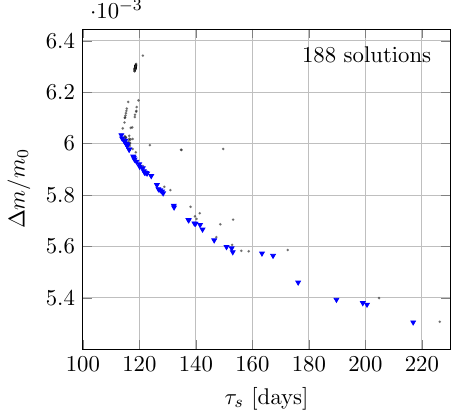}
    \caption{$\alpha=0.7$}
    \label{fig:dvtof_h07}
  \end{subfigure}
  \begin{subfigure}[b]{0.245\textwidth}
    \centering
    \includegraphics[height=3.65cm,keepaspectratio]{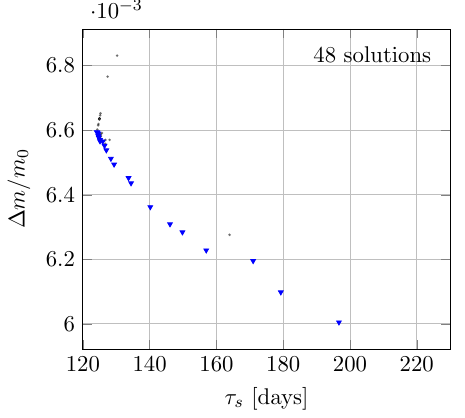}
    \caption{$\alpha=1.0$}
    \label{fig:dvtof_h10}
  \end{subfigure}

  \caption{Pareto fronts of feasible samples (tolerance $e<5\times10^{-5}$ NU) obtained with homotopy and a gradient-based solver.}
  \label{fig:dv-TOF_saturn_SNOPT}
\end{figure}
A more sophisticated alternative is to combine \(\mu\)-homotopy with a gradient-based solver. 
For this approach, we again start from $1000$ feasible solutions to the Europa DRO transfer and use them to initialize the problem at \(\alpha=0.1\) with the same SNOPT settings. 
All feasible solutions found at one homotopy step are then carried forward to the next, while \(\alpha\) is increased by increments of \(0.1\). 
The feasible solutions obtained at each step are shown in the space of the two objectives in Figure~\ref{fig:dv-TOF_saturn_SNOPT}. 
Starting from a fairly dense Pareto front for the original problem, the number of feasible solutions decreases substantially at each step, yielding only $48$ solutions for the final problem and thus a very sparse Pareto front. 
Further reducing the homotopy step size does not improve this behavior. 
With a total runtime of $624$ CPU-hours, this approach is also computationally expensive and we demonstrate in the following sections how MCMC improves these results at a lower computational cost.

The key difficulty is that gradient-based solvers attempt to converge each initialization to a nearby local optimum, whereas the solution distributions evolve nontrivially across the homotopy. 
Solutions for the Titan DRO transfer are not simply direct continuations of corresponding Europa DRO solutions; instead, they can differ qualitatively, for example in the number of throttle switches, so that many homotopy paths terminate early. 
MCMC methods are better suited to this setting because their stochastic proposals allow samples to move between basins of attraction and avoid becoming trapped near local minima.
In addition, even when a particular sample is infeasible at a given iteration, it is not discarded but instead carried forward and further refined in subsequent steps.

\subsection{RWM Performance}
This section presents results for the RWM algorithm, which serves both as a baseline for the more sophisticated MCMC variants and as an initial demonstration of why MCMC is well suited to the present transfer-learning problem.
With $1,920$ initial samples from the baseline diffusion model, the RWM runs require a total of 357 CPU-hours.
Since all components of the algorithm are parallelizable, the computation is distributed across 96 cores.
All MCMC parameters are listed in the left column of Table~\ref{tab:mcmc_params_homotopy_adapted}. 
These parameter values were chosen empirically to provide good performance in the present setting; however, we do not claim that they are optimal, as establishing that would require a dedicated hyperparameter study.

\begin{table}[t!]
  \centering
  \caption{Input parameters for the different MCMC algorithms. 
  Values in brackets indicate adapted parameters for the final iterations with smaller stepsizes.}
  \label{tab:mcmc_params_homotopy_adapted}
  \begin{tabular}{@{}lccc@{}}
    \toprule
    \textbf{Input Parameters} & \textbf{RWM} & \textbf{MALA} & \textbf{HMC} \\
    \midrule
    Objective scaling $\kappa_1$ 
      & $1.0$ 
      & $1.2$ 
      & $1.2$ \\

    Shooting time scaling $\kappa_2$ 
      & $1\times10^{-6}$ 
      & $1\times10^{-6}$ 
      & $1\times10^{-5}$ \\

    Proposal std.\ devs.\ $\boldsymbol{\sigma}_{\lambda}/\boldsymbol{\sigma}_{\mathrm{init}}$ 
      & $0.05$ ($0.002$)
      & $0.02$ ($0.005$)
      & $0.006$ ($0.005$) \\

    Gradient timestep $\epsilon$ 
      & -
      & $2.5$ ($0.1$)
      & $0.5$ ($0.1$) \\

    Leapfrog steps $L$ 
      & -
      & -
      & $3$ ($1$) \\
      
    Scaling factor $\beta$ 
      & $10{,}000$ ($200{,}000$)
      & $10{,}000$ ($200{,}000$)
      & $10{,}000$ ($200{,}000$) \\

    Number of chains $N_{\mathrm{chains}}$ 
      & $1{,}920$ 
      & $1{,}920$ 
      & $1{,}920$ \\

    Number of iterations $M$ 
      & $3{,}000$ 
      & $350$ 
      & $140$ \\

    Burn-in iterations $M_0$ 
      & $2{,}470$ 
      & $270$ 
      & $105$ \\
    \bottomrule
  \end{tabular}
\end{table}
The most important parameter choice for good performance is the diagonal covariance matrix $\boldsymbol{\Sigma}$ in the proposal distribution of Eq.~\eqref{eq: proposal distribution}. 
A naive isotropic covariance performs poorly for this problem because the different components of the costate vector have markedly different scales. Another strategy we tested was to perform a pilot run with fixed covariance in each direction and then choose the proposal covariance as a multiple of the empirical covariance of the generated samples. 
Although this improves performance, the additional pilot run increases the computational cost. 
A better performance is obtained by choosing the proposal covariance $\boldsymbol{\Sigma}=\mathrm{Diag}(\boldsymbol{\sigma_{\lambda}^2})$ such that $\boldsymbol{\sigma}_{\lambda}$ is a scalar multiple of the empirical standard deviation of the initial samples $\boldsymbol{\sigma}_{\mathrm{init}}$.
Here $\boldsymbol{\sigma}_{\lambda}$ is a vector consisting of the standard deviation $\sigma_{\lambda,i}$ for each direction $i$ and $\boldsymbol{\sigma}_{\mathrm{init}}=[0.0468,0.0010,0.0013,0.0353]$.

\begin{figure}[b!]
\centering\includegraphics[width=\textwidth]{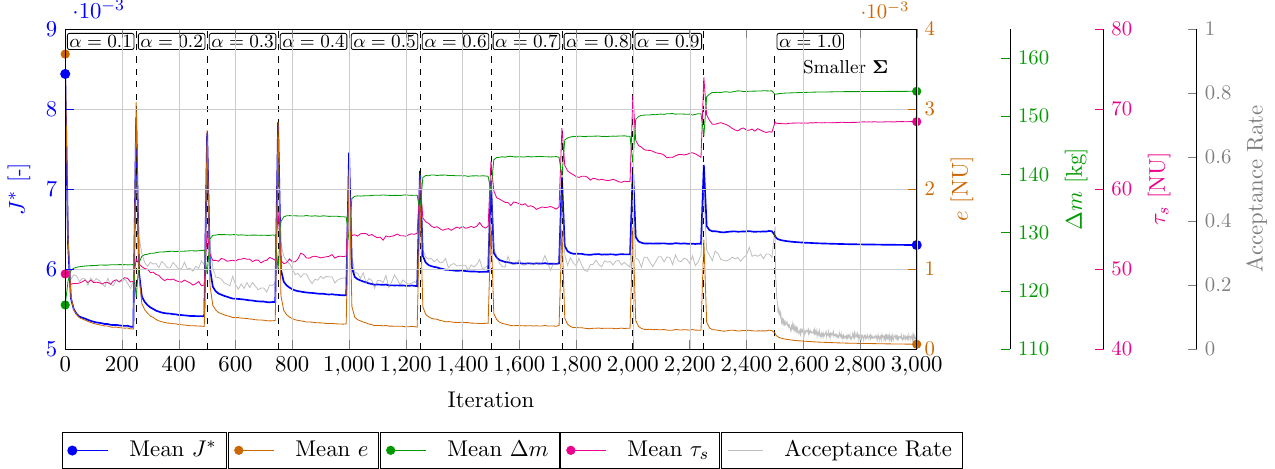}
	\caption{Mean values of the objective function, its three components, and the acceptance rate over the RWM iterations.}
	\label{fig:RWM_iter_plot}
\end{figure}

The objective-function parameter $\kappa_1$ from Eq.~\eqref{eq: objective} was chosen to provide a suitable trade-off between feasibility, that is, low constraint violations, and low objective values.
In contrast, $\kappa_2$ directly governs the trade-off between the two objectives and is selected to target the desired region of the Pareto front.
Finally, the scaling factor $\beta$ from Eq.~\eqref{eq: target distribution} controls the acceptance rate and is chosen to yield values between $20\,\%$ and $40\,\%$, as indicated by the gray curve in Figure~\ref{fig:RWM_iter_plot}.
The MCMC algorithm is run for $250$ iterations at each intermediate system, followed by an additional $500$ iterations in the final system with a reduced proposal covariance.
Combined with an increased $\beta$ value, shown in brackets in Table~\ref{tab:mcmc_params_homotopy_adapted}, these final smaller steps help the algorithm better resolve local minima for the final Saturn-Titan problem. 
Further increasing the number of samples or iterations for this problem only gives marginal improvements.

Across the intermediate homotopy stages, both the mean objective and mean constraint violation drop substantially within each MCMC run. 
This is shown in Figure~\ref{fig:RWM_iter_plot}, which shows the mean values of the objective and its three components over all chains throughout the iterations.
Recurring spikes every 250 iterations mark transitions to the next system (new $\mu$). 
From stage to stage, average fuel consumption and shooting time rise, reflecting the growing energy gap and additional revolutions needed. 
Yet within each stage, after an initial jump, the shooting time trends downward and fuel consumption remains nearly flat.
An additional decrease in $J^*$ is achieved through the decreased proposal covariance and increased $\beta$ in the final stage, as the smaller step size drives down the constraint violations. 
The acceptance rate drops to below $5\,\%$ during these final iterations due to the increased $\beta$ value.
This prevents too many samples with worse objective values from being accepted, as exploration of new local minima is not the primary focus during that stage.
%After discarding the $10\,\%$ samples with the lowest objective value, $31,400$ samples are used to fine-tune the baseline diffusion model using reward-weighted likelihood optimization.
\begin{figure}[b!]
  \centering
  % ---------- Row 1 ----------
  \begin{subfigure}[t]{0.32\textwidth}
    \centering
    \includegraphics[width=\linewidth]{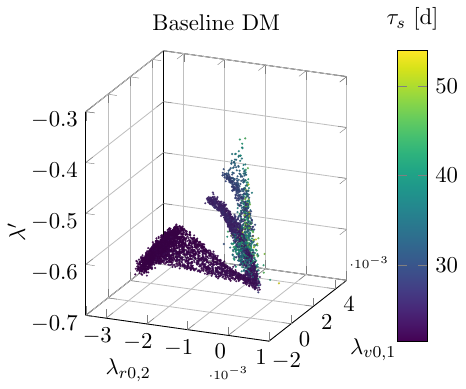}
  \end{subfigure}
  \hfill
  \begin{subfigure}[t]{0.32\textwidth}
    \centering
    \includegraphics[width=\linewidth]{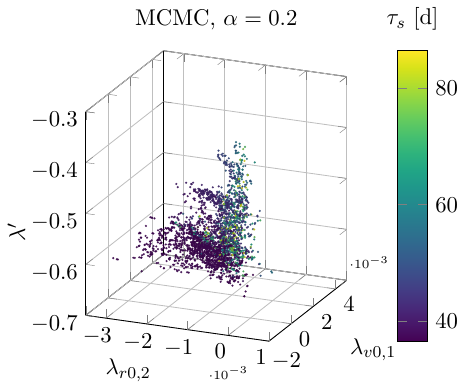}
  \end{subfigure}
  \hfill
  \begin{subfigure}[t]{0.32\textwidth}
    \centering
    \includegraphics[width=\linewidth]{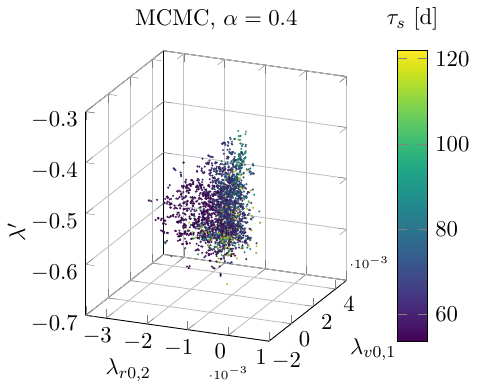}
  \end{subfigure}

  \vspace{0.8em} % vertical gap between rows (tweak as needed)

  % ---------- Row 2 ----------
  \begin{subfigure}[t]{0.32\textwidth}
    \centering
    \includegraphics[width=\linewidth]{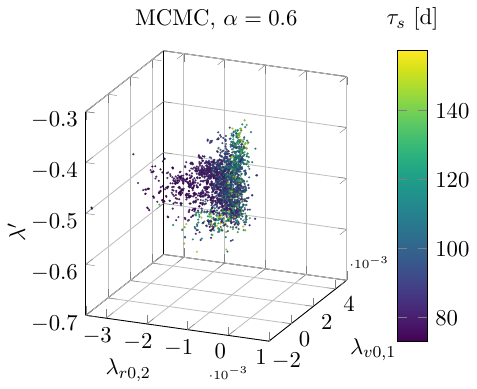}
  \end{subfigure}
  \hfill
  \begin{subfigure}[t]{0.32\textwidth}
    \centering
    \includegraphics[width=\linewidth]{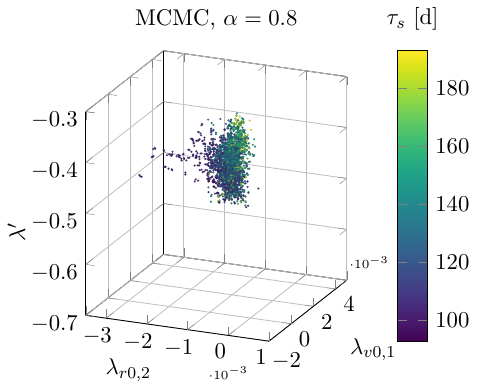}
  \end{subfigure}
  \hfill
  \begin{subfigure}[t]{0.32\textwidth}
    \centering
    \includegraphics[width=\linewidth]{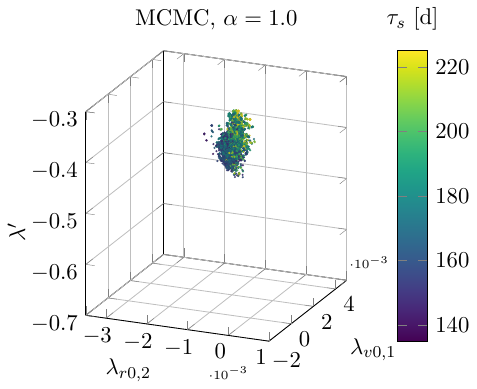}
  \end{subfigure}

  \caption{Comparison of RWM samples in the costate space for different values of $\alpha$.}
  \label{fig:mcmc_grid}
\end{figure}

Figure~\ref{fig:mcmc_grid} illustrates the evolution of the sample cloud in costate space across the intermediate MCMC stages.
The initial distribution for the Europa DRO transfer is captured by the baseline diffusion model. 
It is visualized as a three-dimensional projection of the four-dimensional costate space, obtained by combining two elements of the costate vector as
$\lambda'=0.5895\lambda_{r1}+0.8075\lambda_{v2}$.
This transformation reflects a linear relationship between $\lambda_{r1}$ and $\lambda_{v2}$ that was found in our earlier work~\cite{graebner_JAS}, where the structure of solutions in the costate space for the Europa DRO transfer was analyzed in detail.
Starting from this initial hypersurface structure, the samples diffuse to a new region of the solution space during the homotopy-MCMC procedure.
The considerable distance between the final samples at $\alpha=1$ and most samples from the baseline model explains why going directly from the Europa DRO to the Titan DRO transfer fails.

\subsection{MALA Performance}
Starting from the same \(1{,}920\) samples generated by the baseline diffusion model, we run MALA for a total of \(517\) CPU-hours across \(96\) cores.
All algorithmic parameters are listed in the middle column of Table~\ref{tab:mcmc_params_homotopy_adapted}.
Most MALA parameters are chosen based on those used for RWM.
Different values of the objective-function scaling parameters \(\kappa_1\) and \(\kappa_2\) are selected empirically, as they provide slightly better coverage of the Pareto front for MALA.
In particular, \(\kappa_1\) can be increased from \(1.0\) to \(1.2\) without reducing feasibility, in contrast to the behavior previously observed for RWM.
As before, the proposal covariance \(\boldsymbol{\Sigma}=\mathrm{Diag}(\boldsymbol{\sigma}_{\lambda}^{2})\) is chosen based on the empirical standard deviation of the initial samples, but with a smaller scaling factor, \(\boldsymbol{\sigma}_{\lambda}/\boldsymbol{\sigma}_{\mathrm{init}}=0.02\) instead of \(0.05\).
This smaller covariance is used to avoid increasing the total step size once the gradient-drift term is included.
\begin{figure}[b!]
\centering\includegraphics[width=\textwidth]{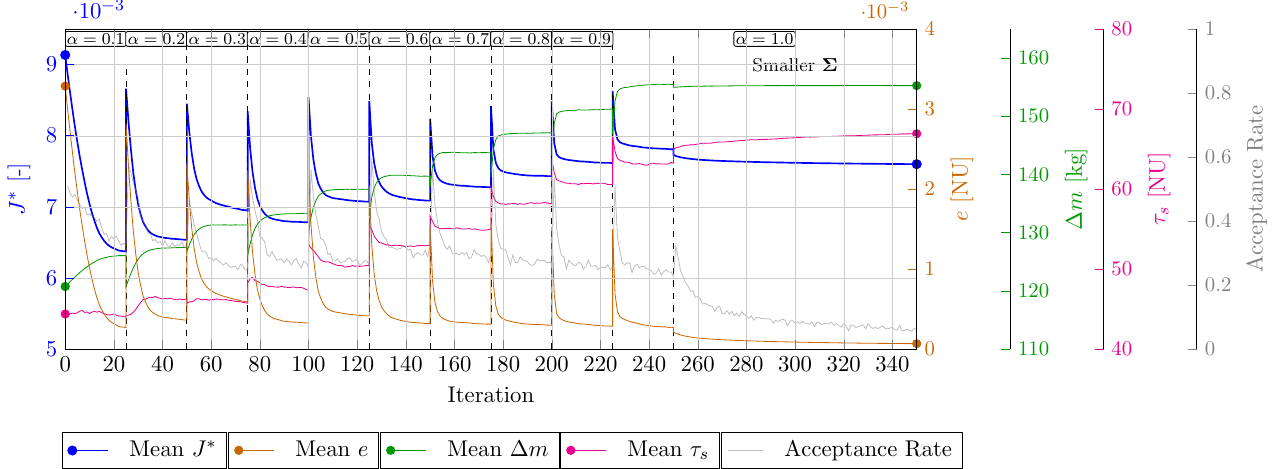}
	\caption{Mean values of the objective function, its three components, and acceptance rate over the MALA iterations.}
	\label{fig:MALA_iter_plot}
\end{figure}
The gradient timestep \(\epsilon\) is chosen so that the gradient step is, on average, approximately equal in magnitude to the random step.
Empirically, this choice yields strong performance, whereas larger values noticeably degrade proposal quality.

MALA produces high-quality proposals that rapidly reduce the mean objective value across all chains.
This behavior is illustrated in Figure~\ref{fig:MALA_iter_plot}, which shows the mean values of the objective and its three components over all chains throughout the iterations.
Only \(25\) iterations are used per homotopy step, followed by an additional \(100\) iterations with smaller step sizes in the final stage, for a total of \(350\) iterations.
This is a substantial reduction compared with the RWM approach and reflects the improved proposal quality obtained by incorporating gradient information.
The improved proposal quality also leads to a higher average acceptance rate, as shown in Figure~\ref{fig:MALA_iter_plot}.

%These final iterations further reduce the mean constraint violations by more effectively capturing local minima through a smaller proposal covariance and gradient step (values shown in brackets in Table~\ref{tab:mcmc_params_homotopy_adapted}).
%When the algorithm switches to the next system in the homotopy scheme, the objective value exhibits a momentary increase caused by larger constraint violations under the modified dynamics. 
%Both the mean constraint violation and the objective value then quickly decrease, accompanied by a rise in mean fuel consumption. 
%This behavior is expected, as a larger $\alpha$ requires a higher $\Delta v$ to achieve feasibility. Following this brief adjustment, the mean fuel consumption remains largely constant.

\subsection{HMC Performance} 
We test the HMC algorithm on the same problem, starting with the identical 1,920 samples from the baseline distribution. 
All algorithmic parameters are provided in the right column of Table~\ref{tab:mcmc_params_homotopy_adapted}.
Because each proposal includes $L$ leapfrog integration steps, each HMC iteration is computationally more expensive than a MALA iteration.
Running the algorithm for a total of $140$ iterations takes $604$ CPU-hours, distributed across $96$ CPUs.
The number of leapfrog integration steps is kept relatively small at $L=3$ to avoid excessive computational cost per iteration. 
For this problem, increasing the total number of iterations provides a better return than allocating the same effort to further increasing $L$, since raising $L$ directly reduces how many iterations can be afforded for a fixed computational budget.
All other parameters are chosen to be similar to the MALA algorithm, with slight modifications chosen empirically based on improved performance.

\begin{figure}[b!]
\centering\includegraphics[width=\textwidth]{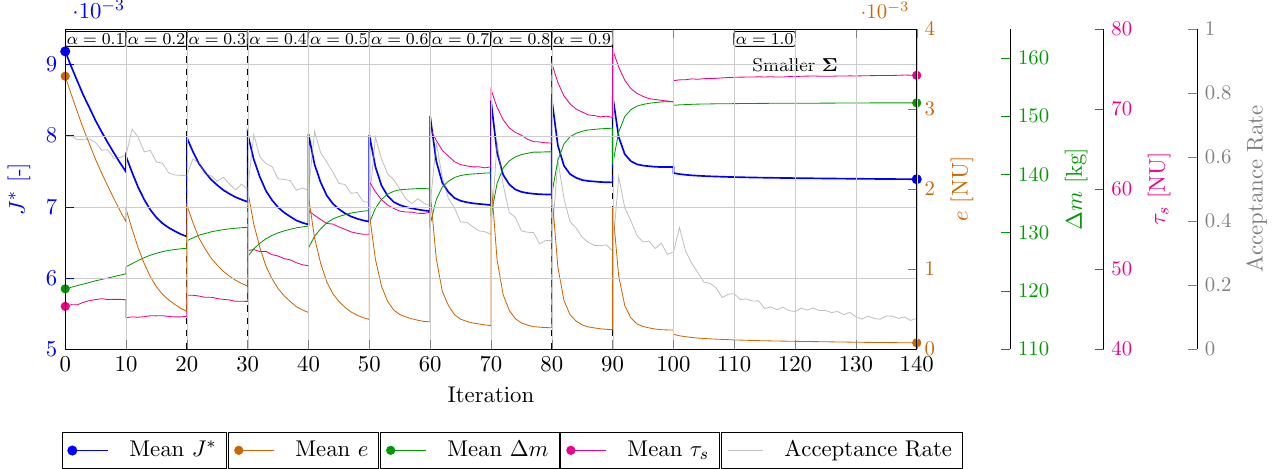}
	\caption{Mean values of the objective function, its three components, and acceptance rate over the HMC iterations.}
	\label{fig:HMC_iter_plot}
\end{figure}

The evolution of the objective function over the iterations is comparable to the MALA case, as shown in Figure~\ref{fig:HMC_iter_plot}.
While only $10$ iterations per homotopy stage are not enough to reduce the mean objective to a level comparable to that achieved in the MALA case with $25$ iterations for $\alpha=0.1$, the subsequent homotopy stages compensate for this, resulting in high-quality samples after the burn-in phase.
Figure~\ref{fig:HMC_iter_plot} also shows that the acceptance rate is consistently higher for HMC than for MALA (Figure~\ref{fig:MALA_iter_plot}).
This is expected, as the leapfrog integration is designed to follow regions of high target density, thereby proposing higher-quality samples that are more likely to be accepted.
In the final stage, an additional $40$ iterations with a decreased step size are added (parameters in brackets in Table~\ref{tab:mcmc_params_homotopy_adapted}).
Because the goal in this final stage is solely to better resolve local minima, rather than to discover new ones, we set $L=1$, effectively reducing the method to MALA.

\subsection{Comparison of MCMC Methods}
We compare the two gradient-based MCMC algorithms to the RWM algorithm, based on the final generated samples for the Titan DRO transfer.
Multiple performance metrics, as well as computational cost, are considered in the comparison.
The results show that incorporating gradient information leads to overall improved performance, even when accounting for the additional computational overhead. 

Feasibility is measured as the percentage of MCMC samples with constraint violations $e<5\times10^{-5}$.
This tolerance matches the value used in our previous work to generate training samples for the baseline model\cite{graebner_JAS}.
Although relatively loose, it is appropriate for the low-fidelity global search considered here, which targets the preliminary mission design phase.
Prior results show that a large fraction of these solutions converge under tighter tolerances when used to warm-start a numerical solver for a related problem\cite{graebner_JAS}.

\begin{figure}[t!]
  \centering

  \begin{subfigure}[b]{0.37\textwidth}
    \centering
    \includegraphics[height=5.0cm,keepaspectratio]{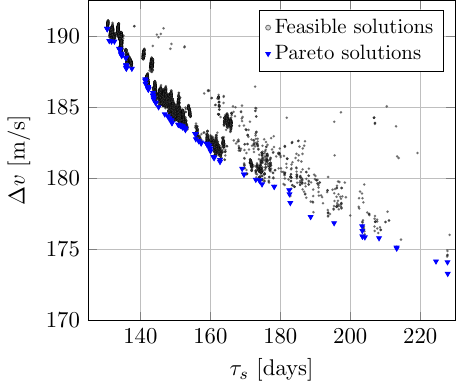}
    \caption{RWM}
    \label{fig:dvtof_rwm}
  \end{subfigure}\hfill
  \begin{subfigure}[b]{0.31\textwidth}
    \centering
    \includegraphics[height=5.0cm,keepaspectratio]{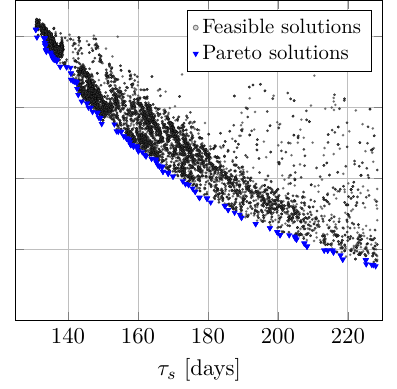}
    \caption{MALA}
    \label{fig:dvtof_mala}
  \end{subfigure}\hfill
  \begin{subfigure}[b]{0.31\textwidth}
    \centering
    \includegraphics[height=5.0cm,keepaspectratio]{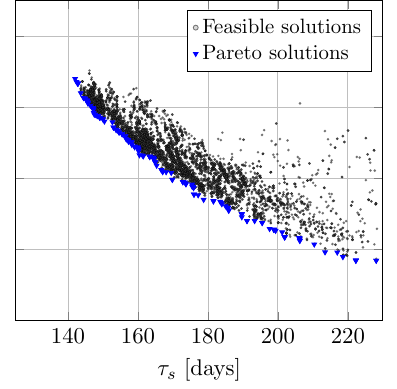}
    \caption{HMC}
    \label{fig:dvtof_hmc}
  \end{subfigure}

  \caption{Feasible samples (tolerance $e<5\times10^{-5}$ NU) from RWM, MALA, and HMC, shown in the $\Delta v$--$\tau_s$ plane.}
  \label{fig:dv-TOF_saturn_comp}
\end{figure}

\begin{table}[b!]
  \centering
  \captionof{table}{Comparison of performance and computational cost for the three MCMC algorithms. Mean and standard deviation (STD) are reported for both objectives; feasibility rate is based on $e<5\times10^{-5}$. All metrics are based on samples collected after the burn-in phase ($15,000$ to $25,000$ per method). Arrows indicate whether larger ($\uparrow$) or smaller ($\downarrow$) values are preferred.}
  \label{tab:MCMC_results}
  \begin{tabular}{@{}lccc@{}}
    \toprule
    & \textbf{RWM} & \textbf{MALA} & \textbf{HMC} \\
    \midrule

    \multicolumn{4}{l}{\textbf{Performance}} \\
    \cmidrule(lr){1-4}
    $\Delta v$ (Mean $\downarrow$ $\pm$ STD $\uparrow$) [m/s] & $185.65\pm2.97$ & $184.95\pm4.15$ & $181.86\pm2.59$ \\
    $\tau_s$ (Mean $\downarrow$ $\pm$ STD $\uparrow$) [days]   & $149.59\pm13.82$ & $158.56\pm24.28$ & $170.83\pm18.41$ \\
    Pareto Front Hypervolume $\uparrow$               & 0.640 & 0.649 & 0.620 \\
    Feasibility rate $\uparrow$                  & 17.34\,\%        & 63.01\,\%        & 37.45\,\% \\[3pt]

    \multicolumn{4}{l}{\textbf{Computational Cost}} \\
    \cmidrule(lr){1-4}
    Runtime [CPU hours]                & $357$ & $517$ & $604$ \\
    Num.\ of function evaluations    & $5,760,000$ & $672,000$ & $652,800$ \\
    Num.\ of gradient evaluations      & $0$ & $672,000$ & $652,800$ \\
    \bottomrule
  \end{tabular}
\end{table}
Table~\ref{tab:MCMC_results} indicates that MALA achieves by far the highest feasibility rate, almost quadrupling the RWM value from $17.34\,\%$ to $63.01\,\%$.
While HMC also increases the feasibility rate relative to RWM, it does not reach the MALA level, likely due to the smaller number of iterations.
Mean values of $\Delta v$ and shooting time $\tau_s$ for all feasible samples indicate that MALA achieves a lower $\Delta v$ than RWM, at the cost of a longer shooting time. 
HMC exhibits an even stronger trend towards lower $\Delta v$ and longer $\tau_s$.
This can also be seen in Figure~\ref{fig:dv-TOF_saturn_comp}, which shows all feasible samples for the three methods in the frame of the two competing objectives. 
The figure demonstrates that HMC misses part of the Pareto front at smaller $\tau_s$.
Even though the larger shooting time scaling factor for HMC ($\kappa_2=1\times10^{-5}$ instead of $\kappa_2=1\times10^{-6}$) helps target the region of small $\tau_s$, the algorithm still excludes portions of that region.
Our experiments indicate that further increasing $\kappa_2$ enables HMC to explore the region of smaller $\tau_s$; however, this comes at the expense of reduced coverage at larger $\tau_s$. 
Consequently, for this specific problem, HMC does not achieve the same breadth of Pareto-front coverage as MALA.
The MALA samples exhibit the greatest diversity in the objective space, observed both visually and quantitatively through the larger standard deviations reported in Table~\ref{tab:MCMC_results}.

Overall, the gradient-based methods achieve a denser Pareto front, particularly at larger values of $\tau_s$.
The quality of a Pareto front is commonly assessed using the hypervolume indicator, which simultaneously reflects convergence toward the true Pareto and the distribution of solutions along it~\cite{797969}. 
Formally, it quantifies the volume of objective space dominated by the Pareto set with respect to a prescribed reference point, with larger values indicating superior front quality. 
The hypervolume values in Table~\ref{tab:MCMC_results} are computed in a normalized objective space, where each objective is linearly scaled to the interval $[0,1]$ using fixed bounds ($\tau_s: [124.08,\,228.41]$ days; $\Delta v: [173.68,\,193.67]$ m/s) to ensure consistency across all methods. 
The dominated volume is evaluated with respect to the normalized reference point $(1.0,\,1.0)$, located at the upper limits of both objectives, ensuring that all Pareto fronts are fully captured.
Consistent with the visual comparison, the larger hypervolume attained by MALA indicates the densest and most evenly spaced Pareto front, compared to the other methods.

Computationally, the per-iteration cost of the gradient-based methods is significantly higher than that of RWM, with HMC being the most expensive.
Computing the gradient involves numerically propagating the STM, which contains quadratically more variables than the state. 
Due to the high sensitivity of the problem, this propagation must be highly accurate and relies on precise interpolation of the switching times.
As a result, despite achieving the reported performance with substantially fewer iterations and function evaluations, MALA still results in a higher overall runtime than RWM. 
Nevertheless, given the significantly improved solution quality, MALA remains the more favorable approach.
Each HMC iteration is even more costly, as it requires $L$ gradient evaluations per iteration, resulting in the longest overall runtime. 
For this problem, this additional computational cost is not justified compared to MALA.
However, for more complex problems in which the initial distribution differs substantially from the target distribution, this conclusion may change.
Finally, a systematic study of how decreasing the integration and interpolation accuracy affects the gradient accuracy could improve the computational cost of gradient-based methods. 

\subsection{Comparison to State-of-the-Art}
% No expert-level knowledge about the physics
% We can explicitly target specific areas of the Pareto front => not possible with ACT
% look at average gap size + variance
Based on the results presented in the previous section, MALA consistently achieves the best overall performance among the considered MCMC methods, combining high feasibility rates with broad Pareto-front coverage for the Titan DRO transfer.
We therefore focus the following comparison on MALA and evaluate its performance against a state-of-the-art global search approach from the literature proposed by Russell~\cite{Russell.2007}. 
This state-of-the-art approach does not employ transfer learning, but instead generates solutions for the Titan DRO problem directly.
It does, however, still rely on prior knowledge of the problem, which is discussed in more detail below. 
The comparison is summarized in Table~\ref{tab:SOA_results} in terms of solution quality, feasibility, and computational cost.
\begin{table}[b!]
  \centering
  \captionof{table}{Comparison of performance and computational cost for the best of our MCMC approaches (MALA) against a state-of-the-art approach from the literature~\cite{Russell.2007}. Mean and standard deviation (STD) are reported for both objectives; feasibility rate is based on $e<5\times10^{-5}$. For MALA, $\Delta v$ and $\tau_s$ metrics are based on feasible samples collected after the burn-in phase. Arrows indicate whether larger ($\uparrow$) or smaller ($\downarrow$) values are preferred. }
  \label{tab:SOA_results}
  \begin{tabular}{@{}lccc@{}}
    \toprule
    & \textbf{\shortstack{Russell ACT \\ (original ranges)}} 
    & \textbf{\shortstack{Russell ACT \\ (problem-adapted ranges)}} 
    & \textbf{\shortstack{MALA \\ (this work)}} \\
    \midrule

    \multicolumn{4}{l}{\textbf{Performance}} \\
    \cmidrule(lr){1-4}
    $\Delta v$ (Mean $\downarrow$ $\pm$ STD $\uparrow$) [m/s]  & $192.86\pm1.82$ & $188.56\pm1.90$ & $184.95\pm4.15$ \\
    $\tau_s$ (Mean  $\downarrow$$\pm$ STD $\uparrow$) [days]   & $127.23\pm9.51$ & $138.67\pm8.02$ & $158.56\pm24.28$ \\
    Hypervolume $\uparrow$               & 0.575 & 0.619 & 0.649 \\
    Number of feasible solutions $\uparrow$     & $7,109$ & $9,525$ & $13,347$ \\
    Feasibility rate $\uparrow$                   & 2.17\,\% & 3.26\,\% & 63.01\,\%  \\[3pt]

    \multicolumn{4}{l}{\textbf{Computational Cost}} \\
    \cmidrule(lr){1-4}
    Runtime [CPU hours]                & $517$ & $517$ & $517$ \\
    Num.\ of function evaluations      & $11,195,061$ & $12,786,912$ & $672,000$ \\
    Num.\ of gradient evaluations      & $2,211,765$ & $1,846,875$ & $672,000$ \\
    \bottomrule
  \end{tabular}
\end{table}

The method of Russell~\cite{Russell.2007} employs adjoint control transformations~\cite{Dixon.1981} (ACT) to generate initial costates, which are subsequently used to initialize SNOPT. 
A forward shooting transcription is applied to solve the resulting two-point boundary value problem associated with the indirect method, with SNOPT operated in feasibility mode.
The ACT method introduces a mapping from the in-plane thrust angle $\delta$, the switching function $S$, and their time derivatives to the position and velocity costates:
$(\delta,\dot{\delta},S,\dot{S})\rightarrow(\boldsymbol{\lambda}_r,\boldsymbol{\lambda}_v)$.
These control parameters are sampled uniformly from prescribed ranges. 
Two configurations are considered. 
In the first, the ranges are taken directly from those identified by Russell for the original Europa DRO transfer ($\delta\in\pi+[-0.05,0.05]$, $\dot{\delta}\in[-0.05,0.05]$, $S\in[0.01,0.21]$, and $\dot{S}\in[-0.002,0.0]$). 
In the second, starting from these original ranges, the bounds were iteratively adjusted to better capture Pareto-optimal solutions for the Titan DRO transfer ($\delta\in\pi+[-0.0036,0.009]$, $\dot{\delta}\in[-0.01,0.01]$, $S\in[0.0,0.035]$, and $\dot{S}\in[-0.0012,0.0]$).
In line with Russell’s approach, candidate initial costates generated by the ACT method are filtered through a single propagation of the combined state equations (Eq.~\eqref{Equation: Combined state equations}), and only those satisfying a global search tolerance of $5\times10^{-4}$ are used to initialize the numerical optimization.

\begin{figure}[b!]
  \centering
    
  \begin{subfigure}[b]{0.37\textwidth}
    \centering
    \includegraphics[height=5.0cm,keepaspectratio]{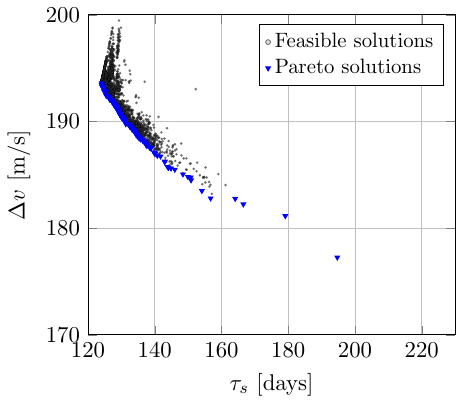}
    \caption{Russell ACT (original ranges)}
    \label{fig:dvtof_act}
  \end{subfigure}\hfill
  \begin{subfigure}[b]{0.31\textwidth}
    \centering
    \includegraphics[height=5.0cm,keepaspectratio]{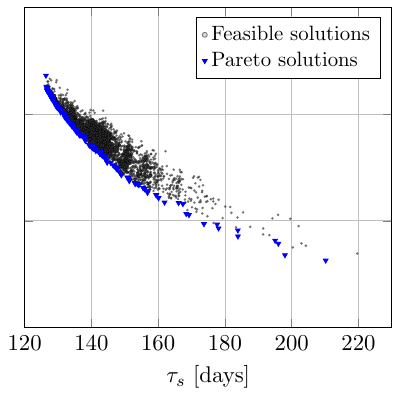}
    \caption{Russell ACT (adapted ranges)}
    \label{fig:dvtof_finetuned}
  \end{subfigure}\hfill
  \begin{subfigure}[b]{0.31\textwidth}
    \centering
    \includegraphics[height=5.0cm,keepaspectratio]{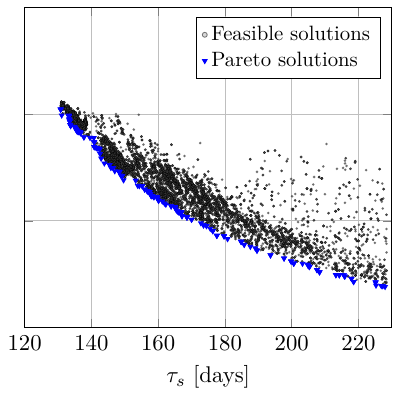}
    \caption{MALA (this work)}
    \label{fig:dvtof_mala_comp}
  \end{subfigure}

  \caption{Feasible samples (tolerance $e<5\times10^{-5}$ NU) from different approaches, shown in the $\Delta v$--$\tau_s$ plane.}
  \label{fig:dv-TOF_saturn_comp_SOA}
\end{figure}

For an identical runtime, our approach outperforms the state-of-the-art in both the number of generated solutions and solution quality, as summarized in Table~\ref{tab:SOA_results}.
While ACT with problem-adapted ranges yields a higher number of feasible solutions and a higher feasibility rate than the original ranges, these numbers remain low relative to the overall runtime, underscoring the difficulty of the problem. 
The poor performance of ACT when using the original ranges from the Jupiter-Europa transfer further indicates that the DRO transfer in the Saturn–Titan system exhibits fundamentally different characteristics from the Jupiter–Europa case.
For the same runtime, MALA produces approximately $40\,\%$ more feasible solutions than ACT with problem-adapted ranges.
Moreover, the final MALA samples exhibit greater diversity within the analyzed shooting time range.
This behavior is illustrated in Figure~\ref{fig:dv-TOF_saturn_comp_SOA}, which shows all feasible samples for the three cases in the space of the two competing objectives.
In contrast, both ACT variants display pronounced clustering in this space, a trend that is also reflected in the smaller standard deviations of both objectives reported in Table~\ref{tab:SOA_results}.

%\begin{table}[b!]
%\centering
%\caption{Comparison of Pareto front quality metrics across the two state-of-the-art approaches and our MALA approach. In addition, the right column shows results obtained after training a DM on the MALA data and warm-starting SNOPT with DM samples.
%All distances are computed in fixed-range normalized objective space. Arrows indicate whether larger ($\uparrow$) or smaller ($\downarrow$) values are preferred.}
%\label{tab:pareto_comparison}
%\begin{tabular}{lcccc}
%\toprule
%\textbf{Metric}
%& \textbf{\shortstack{Russell ACT \\ (original ranges)}} 
%& \textbf{\shortstack{Russell ACT \\ (problem-adapted ranges)}} 
%& \textbf{\shortstack{MALA \\ (this work)}} 
%& \textbf{\shortstack{Fine-tuned DM\\ and SNOPT}} \\
%\midrule
%Pareto solutions $\uparrow$          & 134 & 172 & 107 & 316 \\

%Mean NN distance $\downarrow$        & 0.0067 & 0.0052 & 0.0093 & 0.0034 \\

%Std. NN distance $\downarrow$        & 0.025 & 0.0118 & 0.0070 & 0.0051 \\

%Hypervolume $\uparrow$               & 0.575 & 0.619 & 0.649 & 0.660 \\

%Hypervolume (3$\sigma$ filtered) $\uparrow$ 
%                                    & 0.486 & 0.599 & 0.648 & 0.654 \\

%Hypervolume (1$\sigma$ filtered) $\uparrow$ 
%                                    & 0.410 & 0.582 & 0.638 & 0.645 \\
%\bottomrule
%\end{tabular}
%\end{table}
MALA achieves a superior Pareto front relative to the state-of-the-art approaches.
This is observed both visually in Figure~\ref{fig:dv-TOF_saturn_comp_SOA} and quantitatively through a larger hypervolume indicator in Table~\ref{tab:SOA_results}.
The hypervolume indicator is calculated in the same normalized objective space defined in the previous section. 
%Although the state-of-the-art approaches produce a larger number of Pareto solutions than MALA, these solutions do not evenly span the $\tau_s$ domain. 
%This behavior is further reflected in the nearest-neighbor (NN) distance, for which a high-quality Pareto front is characterized by both a low mean and a low standard deviation.
%While the state-of-the-art approaches attain a lower mean NN distance than MALA, their standard deviation is substantially larger, indicating pronounced clustering.
%This observation underscores the relevance of the hypervolume indicator in this context, as it captures the coverage, density, and diversity of the front.
%Accordingly, the hypervolume results demonstrate that MALA achieves the strongest overall Pareto front coverage.
%After filtering out points that fall outside the 1$\sigma$ and 3$\sigma$ ranges of the nearest-neighbor distance distribution, the superior performance of MALA becomes even more pronounced.
MALA fails to recover a small portion of the Pareto front corresponding to short flight times.
This behavior is likely attributable to the steepness of the Pareto front in this region, where achieving shorter times of flight requires a disproportionately large increase in $\Delta v$, rendering this trade-off unfavorable for the algorithm.
If such solutions are of interest for a particular mission design, they can be explicitly targeted by increasing $\kappa_2$.

While all three approaches considered in this section leverage prior knowledge of solutions to a related problem, the Russell ACT approach with adapted ranges requires additional manual tuning of the control-variable ranges.
This tuning process is labor-intensive, involves multiple test runs, and demands an expert-level understanding of the underlying problem physics.
By contrast, no such manual tuning is required for the MALA-based approach, and this additional workload is not reflected in the direct performance comparison.
It should also be noted that the comparison in Table~\ref{tab:SOA_results} considers only performance on the final Saturn–Titan problem. 
In the proposed transfer-learning framework, however, MALA also generates solution data for all intermediate $\mu$ values encountered along the homotopy. 
These additional samples are subsequently used to fine-tune the diffusion model across the full parameter range. 
If the ACT-based approach were extended to generate solutions for all intermediate systems as well, its total computational cost would increase substantially beyond the final-stage runtime reported here.
%Instead, the $\kappa_1$ and $\kappa_2$ parameters in the proposed MCMC framework provide a direct and intuitive mechanism for targeting specific regions of the Pareto front.
%Such explicit control over the targeted objective values is not available in the ACT approach, which only permits indirect and limited targeting through the specification of control-variable ranges.

Beyond the direct comparison above, the generated MALA samples also form the basis for diffusion-model fine-tuning, which enables subsequent generation of new solutions, as shown in the following section.

\subsection{Diffusion-Model Fine-Tuning}
%14min33sec GPU training
%9.24\% after DM training
%$\Delta v$ (Mean $\pm$ STD) [m/s]: $180.57\pm3.06$ 
%$\tau_s$ (Mean $\pm$ STD) [days]: $186.31\pm21.61$
%$21,182$ final samples to train the DM
%Planning to include the results from actually fine-tuning the DM with MALA samples here. 
%Potentially some analysis of Pareto optimal solutions and the distribution learnt by the DM.
As the final step of the framework, the costate samples generated by MALA, together with their reward values, are used to fine-tune the baseline diffusion model according to Eq.~\eqref{eq:updated loss}.  
Across all \(10\) homotopy stages, MALA produces a total of \(96{,}850\) samples.  
The worst \(10\,\%\) in terms of objective value \(J^*\) are discarded, and the remaining samples are normalized and used for training conditioned on $\alpha$.  
The diffusion-model architecture and hyperparameters are identical to those used in our previous work, where the training procedure is described in more detail~\cite{graebner_JAS}.  
Using a GPU, model training takes approximately \(90\) minutes, and generating \(50{,}000\) samples requires about \(13\) minutes.  
The fine-tuned diffusion model is then evaluated by generating samples for the Titan DRO transfer corresponding to \(\alpha=1.0\).

\begin{figure}[b!]
  \centering
  % --- one height for all three ---
  \newlength{\triFigH}
  \setlength{\triFigH}{5.0cm} % <- adjust once to taste

  % --- choose widths that sum to \textwidth (adjust as needed) ---
  \begin{minipage}[c]{0.31\textwidth}
    \centering
    \includegraphics[height=\triFigH,keepaspectratio]{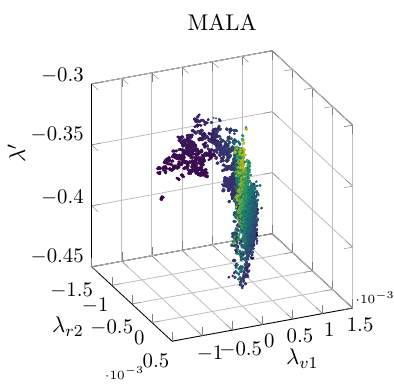}
  \end{minipage}\hfill
  \begin{minipage}[c]{0.31\textwidth}
    \centering
    \includegraphics[height=\triFigH,keepaspectratio]{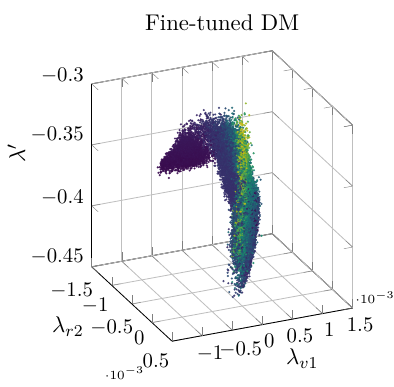}
  \end{minipage}\hfill
  \begin{minipage}[c]{0.37\textwidth}
    \centering
\includegraphics[height=\triFigH,keepaspectratio]{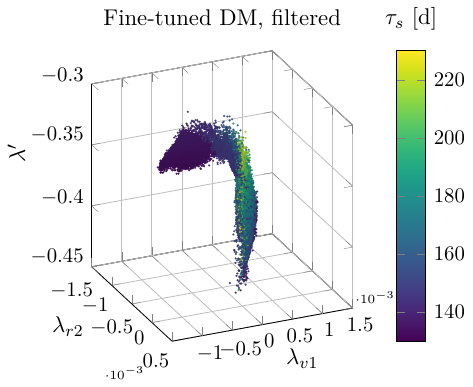}
  \end{minipage}
  \caption{The $19,000$ final MALA samples (left), $50,000$ samples from the fine-tuned DM (middle), and a subset of $21,000$ feasible DM samples (right) with $e<5\times10^{-5}$ displayed in the costate space for $\alpha=1.0$}
  \label{fig:costate_saturn_comp}
\end{figure}

Figure~\ref{fig:costate_saturn_comp} shows a three-dimensional projection of the four-dimensional costate space obtained by combining two elements of the costate vector according to 
$\lambda'=\cos{(0.7766)}\lambda_{r1}+\sin{(0.7766)}\lambda_{v2}$. 
This transformed costate variable is introduced because $\lambda_{r1}$ and $\lambda_{v2}$ exhibit a strong linear relationship with slope $\tan (0.7766)$.
A similar relationship, with a different slope, was observed and analyzed for the related Europa DRO transfer in our previous work~\cite{graebner_JAS}.
The fine-tuned diffusion model learns an approximation of the target distribution and can therefore generate new high-quality samples.  
The final MALA samples for \(\alpha=1.0\), shown on the left of Figure~\ref{fig:costate_saturn_comp}, are approximately distributed according to the target distribution but still exhibit visible gaps.  
The middle and right panels show all samples generated by the fine-tuned diffusion model and the feasible subset of those samples, respectively.  
These plots indicate that sampling from the diffusion model reveals additional solutions and fills regions of the solution space that were not reached by MALA within the allotted number of iterations.

\begin{figure}[t!]
  \centering
  \begin{subfigure}[b]{0.48\textwidth}
      \centering
      \includegraphics[width=\linewidth]{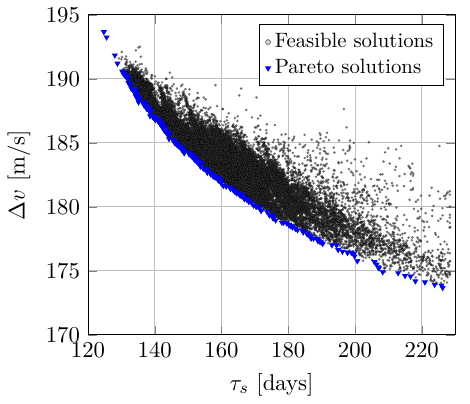}
      %\caption{Fine-tuned DM + SNOPT}
      \label{fig:dvtof_finetuned_1}
  \end{subfigure}
  \hfill
  \begin{subfigure}[b]{0.48\textwidth}
      \centering
      \includegraphics[width=\linewidth]{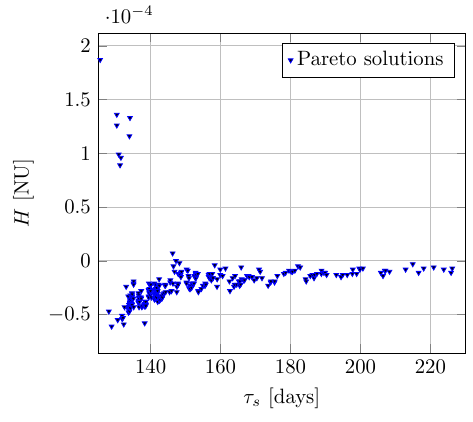}
      %\caption{Fine-tuned DM + SNOPT}
      \label{fig:dvtof_finetuned_2}
  \end{subfigure}

  \caption{Pareto front in the objective frame obtained by warm-starting SNOPT with samples from the fine-tuned diffusion model for $\alpha=1.0$ (left). The Hamiltonian values of the Pareto-optimal samples are shown on the right.}
  \label{fig:dvtof_finetuned_pair}

\end{figure}

Samples from the fine-tuned diffusion model can also be used to warm-start SNOPT, producing the improved Pareto front shown in the left panel of Figure~\ref{fig:dvtof_finetuned_pair}.  
The additional exploration provided by the diffusion model and the solver uncovers new solutions that fill previously observed gaps in the MALA training data. 
%The superior quality of this Pareto front is reflected in the right column of Table~\ref{tab:pareto_comparison}, where it attains the best values across all metrics.
%It should be noted that these gains come at the expense of additional computational effort, while the remaining methods in Table~\ref{tab:pareto_comparison} are compared under equal computational cost.
A subset of the Pareto-optimal samples satisfies, or nearly satisfies, the transversality condition \(H=0\) from Eq.~\eqref{Equation: lam_m=-k}, as illustrated in the right panel of Figure~\ref{fig:dvtof_finetuned_pair}.  
This suggests that these solutions are candidates for local optimality with respect to the single-objective problem of minimizing fuel consumption.

Although we do not evaluate the fine-tuned diffusion model on interpolation to unseen values of \(\alpha\), the presented results still highlight the two main reasons for including the final diffusion-model fine-tuning step in the framework.  
First, learning an approximation of the target distribution enables the generation of an effectively unlimited number of high-quality samples.  
Second, the fine-tuned model uncovers additional solutions by filling regions of the solution space that were not reached by MALA within the available iterations.

% Ideas for additional studies:
% Detailed study of kappa_1 and kappa_2 variations
% Condition on the objective values to target Pareto optimality 

\section{Limitations and Future Work}
\label{sec: limitations}
The proposed method of using MCMC for multiobjective indirect trajectory optimization is primarily motivated by the transfer-learning setting considered in this work, where solution data are already available for at least one mission-parameter value, and the goal is to generate additional training data for a diffusion model. 
In principle, the method could also be applied more broadly by initializing the MCMC procedure from samples drawn from a simpler reference distribution rather than from existing solution data. 
In that case, however, substantially more effort may be required to design an effective proposal mechanism, and the computational advantage relative to a standard gradient-based numerical solver would need to be re-evaluated.

Future work will extend the framework to training-data generation conditioned on other mission parameters. 
Using the mass parameter as the conditioning variable served as an initial demonstration, and the same methodology in principle carries over to other parameters such as maximum thrust magnitude or boundary conditions. 
With minor changes, the framework is also applicable to conditioning on multiple or higher-dimensional mission parameters.

Since the present work focuses primarily on the MCMC component of the framework, the final fine-tuned diffusion model is not evaluated extensively. 
In particular, a detailed study of its interpolation capability for intermediate values of the mission parameter \(\alpha\) is beyond the scope of the current effort and left for future work. 
However, without the transfer-learning extension, we already demonstrated this interpolation capability in our previous work for diffusion models trained directly on datasets spanning multiple parameter values~\cite{graebner_JAS}.

At present, the framework has only been tested on a planar transfer, corresponding to sampling in a four-dimensional costate space. 
The methodology extends directly to spatial transfers with a six-dimensional costate domain, but the resulting impact on sampling efficiency and overall computational cost remains to be quantified.

\section{Conclusion}
\label{sec: conclusion}

This work introduced a transfer-learning framework for indirect multiobjective trajectory optimization that combines parameter homotopy and MCMC sampling for diffusion-model fine-tuning. 
The framework was developed to address a central limitation of earlier diffusion-model approaches: the high cost of generating training data for new mission-parameter values. 
As a test case, we considered an LT transfer problem in the CR3BP with the mass parameter \(\mu\) as the varying mission parameter, and generated data across the full range between the Jupiter-Europa and Saturn-Titan systems. 
Rather than solving each new problem independently, the proposed method detailed in this manuscript transfers information across parameter values by performing homotopy over solution distributions and using MCMC to generate new training samples efficiently at all intermediate systems.

The results show that this is a challenging setting in which transferring solutions across parameter values with a standard gradient-based solver performs poorly.
Among the MCMC methods considered, incorporating gradient information substantially improves performance by increasing proposal quality and reducing the number of iterations required to obtain high-quality samples. 
Among the three algorithms tested, MALA achieved the strongest overall performance, providing the best balance of feasibility, solution quality, Pareto-front coverage, and computational cost.
The solutions generated by MALA for the final Saturn-Titan DRO transfer were also compared with a state-of-the-art approach based on adjoint control transformations and a numerical solver that solves the target problem directly. 
Even though this comparison considered only the final target problem and did not account for the additional solutions generated by MALA at intermediate $\mu$ values, MALA still produced more feasible solutions and achieved broader Pareto-front coverage under the same computational budget.

Fine-tuning the diffusion model on the generated samples allows it to learn a global representation of the target distribution conditioned on the mission parameter.
The fine-tuned model was able to generate additional high-quality samples that filled gaps in the solution space and led to a denser Pareto front after warm-starting the numerical solver. 
Taken together, these results demonstrate that MCMC is well suited for transfer learning in indirect multiobjective trajectory optimization and that combining it with diffusion-model fine-tuning provides an effective way to generate solution data across a continuous range of mission-parameter values.

%Future work will extend these methods to more complex mission architectures, incorporate additional conditioning parameters, and further improve computational efficiency.
%comment on different DM architectures, or backpropagation

\section{Acknowledgement}

Simulations were performed on computational resources managed and supported by Princeton Research Computing, a consortium of groups including the Princeton Institute for Computational Science and Engineering (PICSciE) and the Office of Information Technology’s High-Performance Computing Center and Visualization Laboratory at Princeton University.
The first author acknowledges support for this work through the Princeton MAE Summerfield Second Year Fellowship.
The second author would like to acknowledge partial support for this effort from the Princeton Laboratory for Artificial Intelligence’s [PLI/AI\textsuperscript{2}/NAM] initiative.

\section{Conflict of Interest}

The authors state that there is no conflict of interest.

\bibliographystyle{AAS_publication}   % Number the references.
\bibliography{references}   % Use references.bib to resolve the labels.

\end{document}